\documentclass{aa}  

\usepackage{graphicx}
\usepackage{txfonts}
\usepackage{hyperref}
\pdfoutput=1
\usepackage{amsfonts}
\usepackage{amsmath}
\usepackage{amssymb}
\usepackage{xcolor}
\usepackage{enumitem}
\usepackage{graphicx}
\usepackage{gensymb}
\usepackage{fancyhdr}
\usepackage{times}
\usepackage{ulem}
\usepackage{subcaption} 
\usepackage{float} 
\usepackage{color}
\usepackage{fancyvrb}
\usepackage{breqn}
\VerbatimFootnotes

\definecolor{purple}{RGB}{76, 0,153}

\newcommand{\be}{\begin{equation}}  \newcommand{\ee}{\end{equation}}
\newcommand{\eqa}[1]{\begin{align}   #1 \end{align}}
\newcommand{\nn}{\nonumber}
\newcommand{\bm}[1]{\mbox{\boldmath{$#1$}}}
\begin{document} 
   \title{KiDS-1000 catalogue:  Weak gravitational lensing shear measurements}
   \author{Benjamin Giblin \inst{1}\thanks{bengib@roe.ac.uk} \and Catherine Heymans\inst{1,2} 
   \and Marika Asgari\inst{1} 
   \and Hendrik Hildebrandt\inst{2}
   \and Henk Hoekstra\inst{3}
   \and Benjamin Joachimi\inst{4}
   \and Arun Kannawadi\inst{5,3}
   \and Konrad Kuijken\inst{3}
   \and Chieh-An Lin\inst{1}
   \and Lance Miller\inst{6}
   \and Tilman Tr\"oster\inst{1}
   \and Jan Luca van den Busch\inst{2}
   \and Angus H. Wright\inst{2}
    \and 
    		 Maciej Bilicki\inst{7} 
    \and Chris Blake\inst{8}
    \and Jelte de Jong\inst{9}
    \and Andrej Dvornik\inst{2}
    \and Thomas Erben\inst{10}
    \and Fedor Getman\inst{11}
    \and Nicola R. Napolitano\inst{12}
    \and Peter Schneider\inst{10}
    \and HuanYuan Shan\inst{13,14}
    \and Edwin Valentijn\inst{9}
          }

   \institute{Institute for Astronomy, University of Edinburgh, Royal Observatory, Blackford Hill, Edinburgh, EH9 3HJ, UK 
   \and
   Ruhr-Universit{\"a}t Bochum, Astronomisches Institut, German Centre for Cosmological Lensing (GCCL), Universit{\"a}tsstr. 150, 44801, Bochum, Germany
   \and 
   Leiden Observatory, Leiden University, Niels Bohrweg 2, 2333 CA Leiden, the Netherlands
   \and
   Department of Physics and Astronomy, University College London, Gower Street, London WC1E 6BT, UK
   \and
   Department of Astrophysical Sciences, Princeton University, 4 Ivy Lane, Princeton, NJ 08544, USA
      \and
   Department of Physics, University of Oxford, Denys Wilkinson Building, Keble Road, Oxford OX1 3RH, UK
   \and 
   Center for Theoretical Physics, Polish Academy of Sciences, al. Lotnik\'{o}w 32/46, 02-668, Warsaw, Poland
   \and
   Centre for Astrophysics \& Supercomputing, Swinburne University of Technology, P.O. Box 218, Hawthorn, VIC 3122, Australia 
   	\and 
   	Kapteyn Astronomical Institute, University of Groningen, PO Box 800, 9700 AV Groningen, the Netherlands
      \and
   Argelander-Institut f. Astronomie, Univ. Bonn, Auf dem Huegel 71, D-53121 Bonn, Germany
   \and
   INAF - Astronomical Observatory of Capodimonte, Via Moiariello 16, 80131 Napoli, Italy
   \and
   School of Physics and Astronomy, Sun Yat-sen University, Guangzhou 519082, Zhuhai Campus, P.R. China
   \and
   Shanghai Astronomical Observatory (SHAO), Nandan Road 80, Shanghai 200030, China
   \and
   University of Chinese Academy of Sciences, Beijing 100049, China
    }

  \abstract{
We present weak lensing shear catalogues from the fourth data release of the Kilo-Degree Survey, KiDS-1000, spanning 1006 square degrees of deep and high-resolution imaging.   Our `gold-sample' of galaxies, with well-calibrated photometric redshift distributions, consists of 21 million galaxies with an effective number density of $6.17$ galaxies per square arcminute.  We quantify the accuracy of the spatial, temporal, and flux-dependent point-spread function (PSF) model,  verifying that the model meets our requirements to induce less than a $0.1\sigma$ change in the inferred cosmic shear constraints on the clustering cosmological parameter $S_8 = \sigma_8\sqrt{\Omega_{\rm m}/0.3}$.   Through a series of two-point null-tests, we validate the shear estimates, finding no evidence for significant non-lensing B-mode distortions in the data.   The PSF residuals are detected in the highest-redshift bins, originating from object selection and/or weight bias.   The amplitude is, however, shown to be sufficiently low and within our stringent requirements.   With a shear-ratio null-test, we verify the expected redshift scaling of the galaxy-galaxy lensing signal around luminous red galaxies.   We conclude that the joint KiDS-1000 shear and photometric redshift calibration is sufficiently robust for combined-probe gravitational lensing and spectroscopic clustering analyses. }
 \keywords{gravitational lensing: weak, methods: data analysis, methods: statistical, surveys, cosmology: observations}
   \titlerunning{KiDS-1000 Shear Catalogues}
   \authorrunning{Giblin, Heymans, Asgari \& the KiDS collaboration et al.}
   \maketitle
%
\section{Introduction}
Cosmological information is encoded in the coherent statistical correlations observed between the shapes of background galaxies.  This is a consequence of the weak gravitational lensing of light by foreground large-scale structures.  Combining measurements of the correlations between galaxy shapes, referred to as `cosmic shear', the correlations between the galaxy shapes and the positions of the foreground galaxies, referred to as `galaxy-galaxy lensing', and the correlations between galaxy positions, referred to as `galaxy clustering', provides a powerful set of observables for cosmological parameter inference \citep{hu/jain:2004,joachimi/bridle:2010, zhang/etal:2010}.    The success of this type of study, however, rests on the robustness and accuracy of the core measurement of galaxy shears and 3D positions, with the latter estimated through photometric and/or spectroscopic redshifts \citep[see][and references therein]{mandelbaum:2018}.  

The Kilo-Degree Survey \citep[KiDS,][]{kuijken/etal:2019}, the Dark Energy Survey \citep[DES,][]{drlicawagner/etal:2018}, and the Hyper Suprime-Cam Strategic Program \citep[HSC,][]{aihara/etal:2019} present hundreds to thousands of square-degrees of high-quality deep ground-based multi-band imaging.
Weak lensing analyses of these surveys have already yielded some of the tightest constraints on the clustering parameter $S_8 = \sigma_8 \sqrt{ \Omega_{\rm m}/0.3 }$, where $\sigma_8$ characterises the amplitude of matter fluctuations and $\Omega_{\rm m}$ is the matter density parameter \citep{abbott/etal:2018,troxel/etal:2018,vanuitert/etal:2018,hamana/etal:2019,hikage/etal:2019,hildebrandt/etal:2020,troester/etal:2020}. The success of these investigations builds upon two decades of work from previous generations of weak lensing surveys \citep[see][and references therein]{kilbinger:2015}.

Comparing KiDS with DES and HSC, we recognise that the differences between the survey configurations are largely set by practical considerations associated with instrumentation, resulting in three complementary surveys. 
While DES covers the largest area of sky (almost four times the area of HSC and KiDS), HSC is the deepest, 
with KiDS and DES at roughly the same depth.  In terms of image quality, HSC has the best seeing conditions with a mean seeing of 0.58 arcsec, followed by KiDS with 0.7 arcsec and then DES with $\sim 0.9$ arcsec.  Inspecting the variation of the point-spread function (PSF) across each survey's camera, and seeing variations across each footprint, we conclude that KiDS has the most homogeneous and isotropic PSF, in comparison to DES and HSC. It also has the widest and most extensive matched-depth wavelength coverage, comprising nine bands from $u$ to $K_{\rm s}$.  
In this paper we present the galaxy catalogue of weak lensing shear estimates for the KiDS fourth data release \citep{kuijken/etal:2019}, which totals 1006 square degrees of imaging and hereafter is referred to as KiDS-1000.   

The typical distortion induced by the weak lensing of large-scale structures changes the observed ellipticity of a galaxy by a few percent.   This can be viewed in contrast to the typical distortions induced by the atmosphere, telescope, and camera, encompassed within the PSF, that can alter the observed ellipticity of even a reasonably well-resolved galaxy by a few tens of percent.  Reliable shear estimates therefore require a good understanding of the temporal, spatial, flux, and wavelength variation of the PSF \citep{hoekstra:2004,voigt/etal:2012,massey/etal:2013,antilogus/etal:2014,carlsten/etal:2018} characterised through images of point-source objects.  Efforts to minimise the impact of uncertainties in the PSF model include the installation of cameras that are designed to produce a stable PSF across the field of view with minimal ellipticity \citep{aune/etal:2003,kuijken:2011,flaugher/etal:2015,miyazaki/etal:2018}.  A survey strategy that reserves the best observing conditions for the chosen `lensing' imaging band can also be adopted in order to minimise the PSF size in one of the many multi-band observations \citep[see for example][]{kuijken/etal:2015,aihara/etal:2018}.  This approach is, however, often incompatible with many time domain studies that require a fixed multi-band cadence.  

Shear estimators can be broadly split into two categories: moments-based approaches or model-fitting methods \citep[see the discussion in][]{massey/etal:2007}.  In this paper we adopt the {\it lens}fit likelihood-based 
model-fitting method \citep{miller/etal:2013,fenechconti/etal:2017} which fits a PSF-convolved two-component bulge and disk galaxy model simultaneously to the multiple exposures in the KiDS-1000 $r$-band imaging, returning an ellipticity estimate per galaxy and an associated weight.   This approach is similar to the {\sc im3shape} and {\sc ngmix} model-fitting approaches adopted by DES \citep{zuntz/etal:2013,sheldon:2014,jarvis/etal:2016,zuntz/etal:2018}, differing in the implementation and the choice of galaxy model.  

One of the most important aspects of accurate shear estimation is to quantify the response of the chosen shear estimator to the presence of noise in the images, often referred to as `noise bias'.  Cases of both uncorrelated noise \citep{melchior/viola:2012,refregier/etal:2012}, and correlated noise, for example from the blending of galaxies with unresolved and undetected counterparts \citep{hoekstra/etal:2017,kannawadi/etal:2019,euclidcollab/etal:2019,eckart/etal:2020}, need to be considered.   Noise bias is not the only source of systematic error for shear estimates, however, as during the object detection stage, photometric noise can lead to a preferred orientation in the selection for galaxies aligned with the PSF.  This results in a non-zero mean for the intrinsic ellipticity of the source sample \citep{hirata/seljak:2003,heymans/etal:2006}.    This same effect arises across the full multi-band imaging of the survey which can also lead to photometric redshift selection bias, a bias that is expected to become a significant source of error for 
next-generation surveys \citep{asgari/etal:2019}.     Model-fitting methods are also subject to `model bias', where inconsistencies between the adopted smooth galaxy model and the complex morphology of real galaxies can induce a shear calibration error \citep{voigt/bridle:2010,melchior/etal:2010}.  

There are two main shear calibration approaches to mitigate these sources of bias.   The first is to use the data itself, known as `metacalibration' or `self-calibration'.  In the metacalibration approach, successive shears are applied directly to the data, calibrating the response of the chosen shear estimator at the location of each individual galaxy \citep{sheldon/huff:2017,huff/mandelbaum:2017}.  Self-calibration follows a similar philosophy for model-fitting methods, where the initial best-fit galaxy model, per galaxy, is effectively reinserted into the measurement pipeline.   The difference between the resulting ellipticity measurement and the true input ellipticity is then used as a calibration correction for that galaxy \citep{fenechconti/etal:2017}.   These approaches both mitigate noise bias, with metacalibration also accounting for model bias.  \citet{sheldon/huff:2017} and \citet{sheldon/etal:2019} demonstrate how the metacalibration methodology can also be extended to mitigate object and photometric redshift selection bias. 

The second approach to mitigate shear biases relies on the analysis of realistic pixel-level simulations of the imaging survey \citep[see for example][]{rowe/etal:2015} to determine an average
 shear calibration correction for a galaxy sample 
 \citep{heymans/etal:2006,hoekstra/etal:2015,samuroff/etal:2018,mandelbaum/imsim/etal:2018,kannawadi/etal:2019}.   Provided the image simulations are sufficiently realistic,  the resulting calibration will correct for noise bias including blending, model bias and selection bias.  With realistic multi-band image simulations, photometric redshift selection bias can also be calibrated.

In this paper we adopt a hybrid of both calibration approaches starting with a `self-calibration' stage.   \cite{fenechconti/etal:2017} demonstrated that whilst this approach significantly reduces the amplitude of the noise bias, a percent-level residual remains which is then calibrated, along with the model and selection bias, using image simulations that emulate $r$-band KiDS imaging \citep{kannawadi/etal:2019}.

The conclusion of the shear estimation and calibration analysis follows a succession of `null-tests' to quantify the robustness of shear catalogue to ensure that it is `science-ready'.  The accuracy of the PSF model and correction can be determined through a series of PSF residual size and ellipticity cross-correlation statistics \citep{paulin-henriksson/etal:2008,rowe:2010, jarvis/etal:2016} and through the cross-correlation of the shear estimates and PSF ellipticities \citep[see for example][]{heymans/etal:2012}.   A series of one-point null-tests can be defined to ensure that the average measured shear is uncorrelated with the measured galaxy flux or the properties of the camera, based on the position of the galaxy in the field of view \citep{heymans/etal:2012,jarvis/etal:2016,zuntz/etal:2018,amon/etal:2018,mandelbaum/etal:2018}.  These null-tests can also be extended further to include the properties of the galaxies.  \cite{troxel/etal:2018}, for example, present two-point cosmic shear measurements differenced for a range of galaxy properties such as the measured galaxy size and signal-to-noise.   Given the impact of selection bias when constructing samples from measured, rather than intrinsic galaxy properties, it is often hard to interpret these galaxy-property level null-tests \citep[see the discussion in][]{fenechconti/etal:2017, mandelbaum:2018}.  In this analysis we therefore limit our null-test studies to observables that are clearly uncorrelated with the intrinsic ellipticity of the galaxies.  We also introduce a new two-dimensional (2D) galaxy-galaxy lensing null-test to assess the position dependence of additive biases.

Gravitational lensing only produces detectable E-mode distortions, while unaccounted systematics in the data can produce both E- and B-modes of similar amplitude \citep{crittenden/etal:2002}. We can therefore decompose the 
measured signal into its E- and B-modes, using the B-modes to assess the quality of the data \citep[see for example][]{jarvis/etal:2003}.   There is a range of different statistics that can be used to isolate the 
B-modes in the inferred cosmic shear signal \citep[see the discussion in appendix D6 of][]{hildebrandt/etal:2017}.  In this analysis we adopt the `COSEBIs' statistic which has been demonstrated to act as both a stringent tool to detect B-modes, but also as a diagnostic tool in order to isolate the origin of any B-modes that are detected \citep{asgari/etal:2019}.

In all of the KiDS-1000 weak lensing analyses, the catalogue of shear estimates presented in this paper will be used in conjunction with calibrated photometric redshift distributions \citep{hildebrandt/etal:2020b}.    By assuming a fiducial cosmology, the robustness of any joint shear-redshift catalogue can be assessed by cross-correlating shear measurements separated into tomographic bins with foreground, and background,  galaxy positions \citep{heymans/etal:2012}.  Known as the `shear-ratio test', this combined shear-redshift analysis can provide a final assessment of the input joint shear-redshift catalogue for weak lensing surveys \citep{hildebrandt/etal:2017,prat/etal:2018,hildebrandt/etal:2020, maccrann/etal:2020}.
   
This paper is organised as follows.  We summarise the KiDS-1000 data set and our {\it lens}fit data analysis in Sect.~\ref{sec:Data}.  We document the PSF modelling methodology and validate the accuracy of the PSF model in Sect.~\ref{sec:PSF_Estimation}.  Our suite of shear-catalogue null-tests are presented in Sect.~\ref{sec:Null_tests}, with our joint null-test of the shear and photometric redshift estimates presented in Sect.~\ref{sec:Shear_ratio}.  Finally, we conclude in Sect.~\ref{sec:Conc}.  Unless otherwise specified, calculations and figures use the fiducial set of cosmological parameters specified in Table A.1 of \citet{joachimi/etal:2020}.

\section{Data processing and analysis}
\label{sec:Data}

The Kilo-Degree Survey is a European Southern Observatory multi-band public survey with optical imaging in the $ugri$ bands from the 2.6 m VLT Survey Telescope \cite[VST,][]{capaccioli/schipani:2011, capaccioli/etal:2012}.  These data are combined with overlapping 
near-infrared (NIR) images 
in the $ZYJHK_{\rm s}$ bands from the 4.1 m Visible and Infrared Survey Telescope for Astronomy (VISTA), as part of the VISTA Kilo-degree INfrared Galaxy survey \citep[VIKING,][]{edge/etal:2013}.  The KiDS-1000 analyses focus on the fourth KiDS data release, spanning 1006 deg$^2$ of imaging \citep[ESO-KiDS-DR4,][]{kuijken/etal:2019}.

We extract weak lensing measurements from the deep KiDS $r$-band observations.   These images are taken using the wide-field optical camera OmegaCAM \citep{kuijken:2011}, during dark time and under excellent seeing conditions.  The image scheduler follows the requirement that the PSF full-width half-maximum is below $0.8 \, \rm{arcsec}$, resulting in a mean seeing for the full survey of 0.7 arcsec.  The median limiting $5 \sigma$ point-source magnitude (2 arcsec aperture) is $r=25.02 \pm 0.13$. OmegaCAM features 268 million pixels across 32 CCD detectors, with a 1.013$\times$1.020 deg$^2$ field of view.    Data processing for the $r$-band imaging uses the weak-lensing optimised {\sc THELI} data reduction pipeline \citep{erben/etal:2005, schirmer/etal:2013} to produce, tile by tile, an optimised mean co-addition of the five dithered\footnote{The five-step dither follows a staircase pattern with steps parallel to both the RA and declination axes.   The step sizes are chosen to match the gaps between CCDs (25 arcsec in RA, 85 arcsec in declination), ensuring that all parts of the tile are covered by at least three of the exposures \citep{dejong/etal:2015}.} sub-exposures for object detection, as well as individual unstacked calibrated images for each sub-exposure for the weak lensing shape measurements. 

The multi-band optical $ugri$ imaging is processed through the {\sc Astro-WISE} pipeline to produce co-added images for each filter band with improved multi-band photometric accuracy \citep{mcfarland/etal:2013}.   For the multi-band $ZYJHK_{\rm s}$ imaging we use the `paw print' data reduction from the VISTA Science Archive  \citep{cross/etal:2012}.  Accounting for the area lost to multi-band masks, KiDS-1000 is fully imaged in nine bands with matched depths over a total effective area of 777.4 square degrees\footnote{This effective area is the total survey area that is not excluded by the nine-band composite $ugriZYJHK_{\rm s}$ mask that identifies image defects and overlapping regions, in addition to flagging missing data in one or more of the bands.   This mask is defined on the native OmegaCAM pixel scale of 0.213 arcsec.   We note that the KiDS-1000 analysis is based on the KiDS-ESO data release update DR4.1, correcting for a minor error in the registration of the multi-band masking since the publication of \citet{kuijken/etal:2019}.}.  We refer the reader to \citet{kuijken/etal:2019} and \citet{wright/etal:2019} for further details.    

\subsection{Photometric redshifts and calibration} \label{subsec:photoz}
Photometric redshift point estimates, $z_{\rm B}$, are derived using the Bayesian photometric redshift {\sc BPZ} method \citep{benitez:2000} using the redshift probability prior from \citet{raichoor/etal:2014}.  The complete list of settings adopted for the {\sc BPZ} calculation can be found in table 5 of \citet{kuijken/etal:2019}.   We follow \citet{hildebrandt/etal:2020} in using these point estimates to define five tomographic bins between $0.1 < z_{\rm B} \leq 1.2$ (see Table~\ref{tab:catstats}), where the lower and upper $z_{\rm B}$ limits are based on the reliability of the calibration of these photometric redshifts.  For our primary analysis we estimate the true redshift distributions of the five tomographic bins using a large sample of overlapping spectroscopic redshifts and the self-organising map (SOM) methodology of \citet{wright/etal:2020}.   In this analysis, the mapping from multi-dimensional nine-band KiDS photometry colour-magnitude space to true redshift, is trained for each tomographic bin using a sample of over 25,000 spectroscopic redshifts.    The SOM allows us to locate galaxies from the KiDS photometric sample that lie in any part of colour-magnitude space which is not adequately represented in the spectroscopic sample.   These objects can then be removed to create an accurately calibrated redshift distribution.  We hereafter refer to this SOM-selected photometric sample as the `gold' sample.

The \citet{wright/etal:2020} analysis of a mock survey with KiDS properties, based on the MICE simulation \citep{crocce/etal:2015}, confirms that the SOM approach is more robust than the direct redshift calibration method (DIR) adopted for the KiDS-450 and KV450 cosmic shear analyses \citep{hildebrandt/etal:2017, hildebrandt/etal:2020}\footnote{The DIR method includes the photometric galaxies that the SOM flags as problematic, owing to a lack of representation in the spectroscopic sample.   This results in larger bias and uncertainty for the DIR-calibrated redshift distributions compared to the SOM distributions in the MICE mocks \citep[see][for details]{wright/etal:2020}.   We clarify, however, that the redshift uncertainties adopted in previous KiDS-with-DIR cosmic shear analyses, mitigated this bias \citep{wright/etal:2020b}.}. 
This results in a decrease in the uncertainty on the calibrated mean redshift of each tomographic bin, from $\sigma_{\overline{z}}^{\rm DIR} = [0.039,0.023, 0.026, 0.012, 0.011]$ to $\sigma_{\overline{z}}^{\rm SOM} = [0.010,0.011, 0.012, 0.008, 0.010]$.   We note that this reduction in systematic uncertainty incurs an increase in statistical error, as the SOM-gold selection reduces the effective number density of KiDS-1000 galaxies by 20\%.   This increased statistical error is tolerable, however, as our primary focus is the mitigation of systematic errors.  We therefore use the gold photometric sample throughout this paper \citep[see][for the first application of the SOM redshift calibration method to the cosmic shear analysis of KV450]{wright/etal:2020b}.  We refer the reader to \cite{hildebrandt/etal:2020b} for the details of the KiDS-1000 photometric redshift calibration analysis, which also includes a secondary cross-correlation clustering calibration.

\begin{table*}[t]
  \caption{Properties of the KiDS-1000 `gold' galaxy sample in five tomographic redshift bins. For each
   bin we tabulate: 
   the nominal photometric redshift range, $z_{\rm B}$; the effective number density of the gold photometric sample per square arcminute, $n_{\rm eff}^{\rm gold}$; the measured ellipticity dispersion per component, $\sigma_{\epsilon}$; the median redshift of the bin, $z^{\rm median}_{\rm SOM}$, as defined by the SOM calibration; the mean redshift of the bin,  $\langle z_{\rm SOM} \rangle$; the accuracy and uncertainty on the mean of the redshift calibration, $\delta_z$; and the shear calibration correction, $m$.}
\begin{center}
  \begin{tabular}{ c c c c c c r r }
     Bin & $z_{\rm B}$ range & $n_{\rm eff}^{\rm gold} [{\rm arcmin}^{-2}]$ & $\sigma_\epsilon$ & $z_{\rm SOM}^{\rm median}$ & $\langle z_{\rm SOM} \rangle$ & \multicolumn{1}{c}{$\delta_z$} & \multicolumn{1}{c}{$m$}  \\ \hline 
1 & $0.1 < z_{\rm B} \leq 0.3$ & $ 0.62$ & $ 0.270$ & $  0.2073$ & $  0.2571$ & $  0.0001 \pm   0.0106 $ & $  -0.009 \pm    0.019 $ \\ 
2 & $0.3 < z_{\rm B} \leq 0.5$ & $ 1.18$ & $ 0.258$ & $  0.3590$ & $  0.4027$ & $  0.0021 \pm   0.0113 $ & $  -0.011 \pm    0.020 $ \\ 
3 & $0.5 < z_{\rm B} \leq 0.7$ & $ 1.85$ & $ 0.273$ & $  0.5421$ & $  0.5636$ & $  0.0129 \pm   0.0118 $ & $  -0.015 \pm    0.017 $ \\ 
4 & $0.7 < z_{\rm B} \leq 0.9$ & $ 1.26$ & $ 0.254$ & $  0.7460$ & $  0.7918$ & $  0.0110 \pm   0.0087 $ & $   0.002 \pm    0.012 $ \\ 
5 & $0.9 < z_{\rm B} \leq 1.2$ & $ 1.31$ & $ 0.270$ & $  0.9336$ & $  0.9838$ & $ -0.0060 \pm   0.0097 $ & $   0.007 \pm    0.010 $ \\ 
  \end{tabular}
  \label{tab:catstats}
\end{center}
\end{table*}

\subsection{Weak lensing shear estimates and calibration}
\label{sec:lensfit}
Weak lensing shear estimates, $\epsilon$, and associated weights, $w$, are derived from the simultaneous analysis of the individual $r$-band exposures using the model-fitting {\it lens}fit method \citep{miller/etal:2013, fenechconti/etal:2017}.   For a perfect ellipse with a minor-to-major axis length ratio, $\beta$, and orientation, $\phi$, measured counter clockwise from the horizontal axis, the ellipticity parameters $\epsilon = \epsilon_1 + {\rm i} \epsilon_2$ are given by,
\be
\left(
\begin{array}{c}
\epsilon_1 \\
\epsilon_2
\end{array}
\left)
= \frac{1-\beta}{1+\beta} \,
\right(
\begin{array}{c}
\cos 2\phi \\
\sin 2 \phi
\end{array}
\right) \, .
\label{eqn:e1e2}
\ee
With this ellipticity definition, an estimate of the weak lensing shear, $\gamma$, can be constructed, as $ \langle\epsilon \rangle = \gamma$, to first order \citep{seitz/schneider:1997}.

For this KiDS-1000 analysis, we continue to use the self-calibrating version of {\it lens}fit developed for the KiDS-450 data release, described in \citet{fenechconti/etal:2017} and evaluated in \citet{kannawadi/etal:2019}.   Our PSF modelling strategy is however updated in Sect.~\ref{sec:PSF_Estimation}, to incorporate information from the Gaia mission \citep{gaia/etal:2018}.    With the increase in the number of galaxies in the KiDS-1000 sample,
 we are also able to double the overall resolution, and hence accuracy, of our empirical weight bias correction scheme.  This scheme corrects for correlations between the {\it lens}fit weight, the galaxy ellipticity, and the relative orientation of the galaxy to the PSF.  When aligned in parallel with the PSF, a galaxy will be detected with a higher signal-to-noise, and hence be assigned a higher weight, than when it is aligned perpendicularly to the PSF.  Considering galaxies of fixed isophotal area and signal-to-noise, we also find that galaxies have smaller measurement errors, and hence a higher-than-average weight, at intermediate values of ellipticity.  These correlations naturally lead to additive and multiplicative biases in any weight-averaged shear estimator \citep[see section 2.3 of][]{fenechconti/etal:2017}.

To mitigate the impact of weight bias, we create 250 sub-samples of the full KiDS-1000 galaxy catalogue with 50 quantiles in the absolute local PSF model ellipticity, $|\epsilon^{\rm PSF}|$, and 5 quantiles in PSF model size.  For each sub-sample we map the mean of the {\it lens}fit estimated ellipticity variance as a function of observed galaxy ellipticity, $\epsilon_1$ and $\epsilon_2$, signal-to-noise ratio and isophotal area.  We then correct the weights, which account for the measured ellipticity variance, such that the re-calibrated weights in the sample are not a strong function of the relative PSF-galaxy position angle or of the galaxy ellipticity.    We found that creating sub-samples in terms of the absolute PSF ellipticity, in contrast to individual PSF ellipticity components, as in \citet{hildebrandt/etal:2017}, and then increasing the resolution in the PSF ellipticity sub-sampling by a factor of ten, resulted in a reduction in the full survey-weighted average PSF contamination fraction by a factor of three (see Sect.~\ref{sec:H06sysmod} for further details).

As we find no significant changes in the depth and PSF quality when comparing the KiDS-1000 and KV450 data releases, we continue to use the \citet{kannawadi/etal:2019} image simulations to calibrate the KiDS-1000 shear measurements.  \citet{kannawadi/etal:2019} emulate KiDS imaging using morphological information from Hubble Space Telescope imaging of the COSMOS field \citep{scoville/etal:2007, griffith/etal:2012}.   Adopting the reasonable assumption that the COSMOS galaxy sample is representative of those observed in KiDS, the response of the {\it lens}fit shear estimator to different input shears can be determined, under the KiDS observing conditions.    The emulated COSMOS galaxies are also assigned a redshift, $z_{\rm B}$, obtained from KiDS$+$VIKING photometry of the field such that they exhibit similar noise properties to KiDS-1000.  A shear calibration correction $m$ can then calculated per tomographic bin with the galaxies weighted to match the {\it lens}fit observed size and signal-to-noise distribution of KiDS-1000.

 \citet{kannawadi/etal:2019} show that the derived $m$ value is sensitive to the full joint distribution of galaxy size and ellipticity in the input COSMOS sample. 
 Comparing the fiducial calibration corrections with values derived when erasing the apparent COSMOS size-ellipticity correlations, by randomly assigning galaxy ellipticities, leads to $\sim 2 \%$ differences in the calibration corrections in the first three tomographic bins. 
 For the first two tomographic bins, $\sim 2 \%$ differences were also found when the calibration was derived from the full COSMOS sample, compared to the fiducial calibration derived from the $z_{\rm B}$-binned COSMOS samples.  This effect stems from the correlations that exist between galaxy morphology, physical size and photometric redshift.    Applying a tomographic $z_{\rm B}$ selection to the galaxy sample therefore changes the size-ellipticity correlations and the resulting shear calibration.  

\citet{kannawadi/etal:2019} proposed a conservative approach for cosmic shear analyses, setting a calibration correction uncertainty of $\sigma_m^i = 0.02$ for all $i \in \{1,\dots, 5\}$ tomographic bins. This approach was adopted by \citet{hildebrandt/etal:2020}, assuming 100\% correlation between the calibration errors in the different tomographic bins. We review this proposal in light of the insensitivity of the fourth and fifth tomographic bin to the chosen input COSMOS sample, the fact that $\sim 2 \%$ covers the unlikely and extreme case of zero correlation between galaxy size and ellipticity, and the fact that the uncertainty is included in the cosmological analysis as a Gaussian of width $\sigma_m$ such that more extreme values of $m$ are still permitted within the tails of the Gaussian distribution, albeit down-weighted.    We therefore revise the calibration correction uncertainty used in \citet{hildebrandt/etal:2020}. We adopt the largest difference in the estimated $m$-calibrations between the fiducial, randomised, and non-$z_{\rm B}$ selected analyses from \citet{kannawadi/etal:2019}, with a minimum value of $1\%$.   In this case $\sigma_m = [0.019, 0.020, 0.017, 0.012, 0.010]$ which we assume to be 100\% correlated in our fiducial cosmic shear analysis.  We note that taking the alternative approach of adopting uncorrelated and scaled shear calibration errors \citep[see for example Appendix A of][]{hoyle/etal:2018} did not lead to any significant changes in the resulting KiDS-1000 cosmological parameter constraints \citep{asgari/etal:2020b}. 

It is clear that the application of the SOM-gold selection for the KiDS-1000 galaxies is likely to introduce a new selection effect that needs to be accounted for.     We determine the COSMOS gold-selection from the KiDS photometry of the COSMOS field, meaning that we identify which COSMOS galaxies are poorly represented in our spectroscopic sample.   We then mimic the gold-selection in the fiducial \citet{kannawadi/etal:2019} image simulations by removing these under-represented galaxies from the analysis, before weighting the emulated COSMOS galaxies to match the {\it lens}fit observed size and signal-to-noise distribution of the SOM-gold sample.    We find that the gold-selection changes the $m$ calibration corrections by $(m_{\rm all} - m_{\rm gold}) = 0.008$, in the first and fourth tomographic bins, with negligible changes in the remaining three bins.   We adopt these revised gold calibration corrections, as listed in Table~\ref{tab:catstats}.  

We verify that the gold selection does not significantly impact on the calibration correction uncertainty $\sigma_m$, by determining the gold calibration correction for the fiducial, randomised, and non-$z_{\rm B}$ image simulations. Overall $\sigma_m$ is reduced by $\sim 0.001$ in each redshift bin, a reduction that we choose not to include in our analysis as the impact is so small.   We recognise that a high-accuracy assessment of the impact of a SOM-gold selection requires full multi-band image simulations.    As the changes to the shear calibration introduced by the SOM-gold selection on the single-band \citet{kannawadi/etal:2019} image simulations are within our $\sigma_m$ uncertainty limits however,  we reserve the quantification of any second-order multi-band selection effects to a future analysis.   

Table~\ref{tab:catstats} presents the average statistical properties of the KiDS-1000 shear estimates for each gold sample per tomographic bin\footnote{Prior to photometric redshift selection, the KiDS-1000 weighted galaxy number density is $n_{\rm eff}=8.43$ per square arcminute.   When limiting to the $z_{\rm B}$ range $0.1 < z_{\rm B} \leq 1.2$, the galaxy sample reduces to $n_{\rm eff}=7.66$, with the SOM-gold selection introducing a further reduction with $n_{\rm eff}^{\rm gold}=6.17$.}.  We list the effective number density of galaxies per square 
arcmin, $n_{\rm eff}$, to be taken in 
conjunction with the measured ellipticity dispersion, per component, $\sigma_\epsilon$.   These are two key quantities for the cosmic shear and galaxy-galaxy lensing covariance estimates.   We refer the reader to appendix C3 and C4 of \citet{joachimi/etal:2020} where the estimators for these key quantities are derived for a weighted 
and calibrated ellipticity distribution.   
For an ideal survey with unit shear responsivity estimates (such that $m=0$) these terms reduce to the expressions adopted in previous analyses \citep{heymans/etal:2012}.

\subsection{Blinding strategy}
\label{sec:blinding}
Blinded weak lensing analyses were first advocated and implemented in \citet{kuijken/etal:2015}.  Blinding has since become a standard feature of weak lensing studies 
in order for researchers to remain agnostic towards the key cosmological results, 
until the methodology and data analysis choices have been finalised.   The KiDS collaboration have adopted two approaches to date, blinding the shear measurements and weights \citep{kuijken/etal:2015,hildebrandt/etal:2017}, or the photometric redshift distributions \citep{hildebrandt/etal:2020}.   Each time the KiDS team have analysed three or four versions of the data, where one is the truth, unblinding the results when every stage of the analysis has been completed.   \citet{muir/etal:2020} presents the multi-probe blinding strategy for the DES collaboration whereby data transformations are applied to the observed multi-probe data vector to consistently blind the different observations, in addition to multiplicative shear catalogue level blinding.   \citet{hikage/etal:2019} discuss the two-tiered approach of the HSC collaboration whereby the shear calibration correction $m$ is modified first universally, and then an additional correction is applied by each analysis team to facilitate phased unblinding.  \citet{sellentin:2020} presents an alternative approach whereby the cosmic-shear only or multi-probe covariance matrix is modified.   All these approaches can be tuned to introduce a $\sim \pm 2\sigma$ (or greater) change in the recovered value of $S_8 = \sigma_8 \sqrt{ \Omega_{\rm m}/0.3 }$.

The primary KiDS-1000 science goals are cosmic shear constraints \citep{asgari/etal:2020b} and a joint multi-probe analysis of KiDS-1000 with 
BOSS, the Baryon Oscillation Spectroscopic Survey \citep{heymans/etal:2020}.   
As a multi-probe blinding analysis is invalidated by the already public nature of the BOSS cosmological parameter constraints \citep{sanchez/etal:2017}, we adopt the shear catalogue level blinding strategy of \citet{hildebrandt/etal:2017}. The null-tests presented in this paper were performed whilst the KiDS team was still blind, and we note that the conclusions we draw from all these tests are unchanged for each of the three blinded catalogue versions. For full transparency we record here that in the calculation of the shear calibration correction for the SOM-gold selection, the unblinding of co-author Kannawadi was unavoidable.  The unblinded SOM-gold shear calibration corrections for KV450 \citep{wright/etal:2020b} clarified which KiDS-1000 blind was the truth during the evaluation of the KiDS-1000 SOM-gold shear calibration correction.   This information, however, was not distributed to the rest of the team.

\section{The point-spread function} \label{sec:PSF_Estimation}

	The atmosphere, telescope, and camera all contribute to the overall effective PSF, which is modelled based on the light profile of calibration point sources observed at various positions in the field of view. The model is then interpolated to other locations on the exposure to obtain a model for the full CCD mosaic \citep[see, for example,][]{hoekstra:2004, miller/etal:2013, kitching/etal:2013, lu/etal:2017}.  Galaxy shapes can then be corrected for this effect. 
	
	\subsection{Point source selection}
	\label{sec:stargalsep}
To accurately model the PSF, we require a fully representative sample of stars, with negligible contamination from galaxies.   We follow \citet{kuijken/etal:2015}, by selecting star-like sources, on individual exposures, based on their location in the $(T^{1/2},J^{1/4})$ plane, 
where $T$ is a measure of object size, and $J$ is measure of the concentration of the light distribution.  $T$ is given by the second-order moments of the 2D angular light distribution, $I(\boldsymbol{\theta})$, with $T = Q_{11}+Q_{22}$, and
\be
Q_{ij} = \frac{\int d^2\boldsymbol{\theta} \, w[I(\boldsymbol{\theta})] \, I(\boldsymbol{\theta}) \, \theta_i\,\theta_j }{ \int d^2\boldsymbol{\theta} \, w[I(\boldsymbol{\theta})] \, I(\boldsymbol{\theta}) } \,.
\label{eqn:Moment_Brightness_Profile}
\ee
Here $w[I(\boldsymbol{\theta})]$ is a weighting function, which is typically chosen to encompass the full extent of the light distribution. For the purpose of point source selection, we choose the weight to be an iteratively centred Gaussian of width 0.62 arcseconds.  $J$ is given by the axisymmetric fourth-order moment of the light distribution, with
\be
J = \frac{\int d^2\boldsymbol{\theta} \, w[I(\boldsymbol{\theta})] \, I(\boldsymbol{\theta}) \, |\boldsymbol{\theta}|^4 }{ \int d^2\boldsymbol{\theta} \, w[I(\boldsymbol{\theta})]  \, I(\boldsymbol{\theta}) } \,.
\ee

In previous KiDS analyses, the automatic identification of stars was carried out using a `friends-of-friends' algorithm to locate the compact overdensity of stellar objects in the $(T^{1/2},J^{1/4})$ plane.  In this plane, stars have the smallest sizes and the most concentrated light distributions compared to the full sample of detected objects.  The precise location of the stellar objects in this plane, however, varies from exposure to exposure and chip to chip, dependent on the size and shape of the PSF during the observation.   

Automated star-galaxy separation was found to be successful at selecting a clean sample of stars in 
$\sim 90$\% of the data, 
where the measure of success required the per-chip variance of the residual between the measured and model PSF ellipticity to be less than 0.001.  This value was chosen as it provided a clean divide between catastrophic failures, with residual variance at the level of $>0.0025$, and the tail of the typical noise distribution of the KiDS residual PSF ellipticities.  Automation was found to fail in exposures which contained galaxy clusters where the overdensity of similar-sized galaxies in the cluster resulted in the automated `friends-of-friends' algorithm selecting the galaxy cluster overdensity in the $(T^{1/2},J^{1/4})$ plane, rather than the stellar overdensity.   To remedy this and avoid the manual intervention required for previous KiDS studies, our KiDS-1000 analysis incorporates the DR2 point source catalogue from the Gaia mission \citep{gaia/etal:2018}.   

Objects detected in the KiDS imaging are cross-matched with all Gaia-defined point sources. These are primarily stars, with a low level of contamination from extended sources \citep{arenou/etal:2018}.  This bright catalogue is too sparse to provide an accurate model of the spatially varying PSF for each exposure, but it is sufficient to define the size and concentration of the stellar population in the $(T^{1/2},J^{1/4})$ plane.
For each exposure and CCD, we therefore augment the sample of PSF objects by adding all sources less than three standard deviations away from the mean in the $(T^{1/2},J^{1/4})$ plane. The standard deviations and corresponding directions are defined from a principal component analysis performed on the Gaia-matched KiDS sources on the CCD
\footnote{
The number of Gaia objects per CCD chip varies across the KiDS sky coverage, with an average of 90 per chip in the southern stripe and 120 per chip in the equatorial stripe.
}.  
Adopting this methodology satisfied our requirement 
for low-levels 
of variance in the per-chip residual between the measured and model PSF ellipticity for the full KiDS-1000 sample.

\begin{figure}
\includegraphics[width=0.5\textwidth]{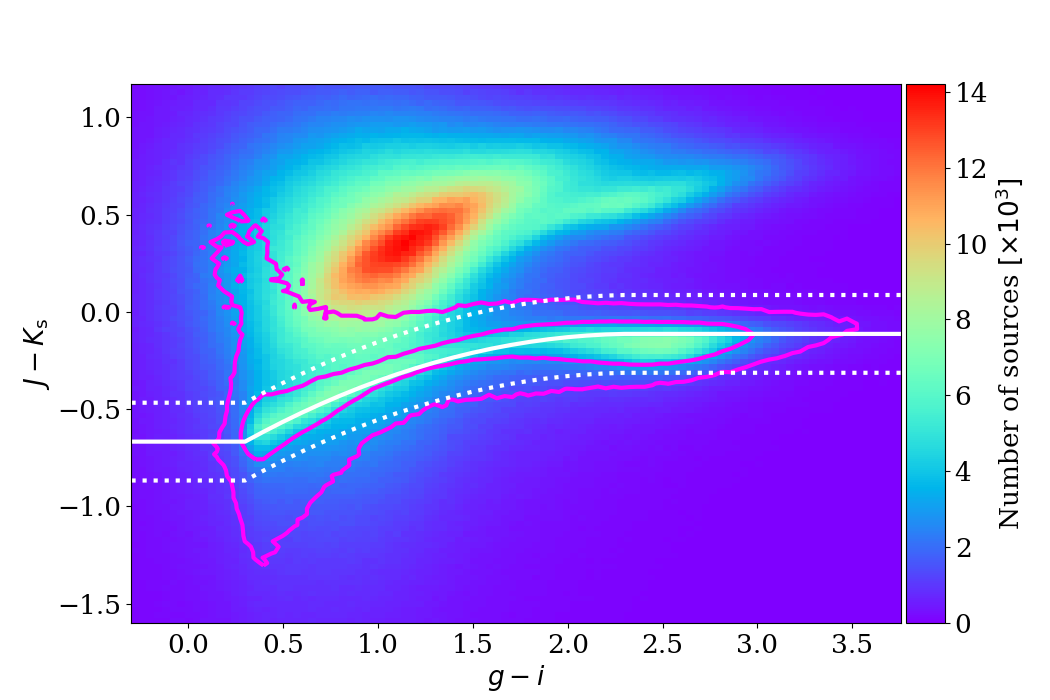} 
\caption{$(J-K_{\rm s}, g-i)$ distribution of the full equatorial KiDS-1000 catalogue (colour map) revealing two distinct populations, compared to the distribution of objects identified as point sources (magenta contours enclosing 68\% and 95\% of the sample).  We find that 3\% of our point-source sample has colours that are inconsistent with the stellar locus and exclusion criteria from \citet[][solid and dotted white lines]{baldry/etal:2010}.    The significant tail of point-source objects extending beyond $(J-K_{\rm s})>0$ and $(g-i)< 0.5$ have colours that are characteristic of quasars, which are combined with the stellar sample in our point-source catalogue.}
\label{fig:star-col-col}
\end{figure}

\citet{baldry/etal:2010} present a star-galaxy separation technique based on object size and NIR-optical colour in the $(J-K_{\rm s}, g-i)$ colour-colour space.  Using spectroscopy from the Galaxy And Mass Assembly survey \citep[GAMA,][]{driver/etal:2011}, they demonstrate that their selection criteria result in a galaxy selection that is 99.9\% complete.    Figure~\ref{fig:star-col-col} compares the $(J-K_{\rm s}, g-i)$ distribution of the full sample of objects in the equatorial KiDS-1000 region (shown as a colour scale) to the distribution of our point-source sample (shown as magenta contours enclosing 68\% and 95\% of the sample).   We see that this combination of NIR and optical colours defines two distinct populations,  with very similar results found for the southern KiDS-1000 region, albeit with a lower stellar density resulting from the increased distance from the Galactic equator.    The stellar locus, from equation 2 of \citet{baldry/etal:2010}, and the GAMA-defined exclusion criteria are shown as solid and dashed white lines.   We find that the majority of the widening of the contours seen around $(g-i)\sim 0.5$ derives from an increase in the average $(J-K_{\rm s})$ photometric error, $\sigma_{(J-K_{\rm s})}$, 
and 
that only 3\% of our point-source sample 
have 
colours that are inconsistent, at more than $3\sigma_{(J-K_{\rm s})}$, with a \citet{baldry/etal:2010} defined stellar population.    We note that the significant tail of point-source objects extending beyond $(J-K_{\rm s})>0$ and $(g-i)< 0.5$ have colours that are characteristic of quasars \citep[see for example figure 6 in][]{baldry/etal:2010}.   As high-redshift quasars are suitable point sources to include in our PSF model\footnote{The wavelength range of the OmegaCAM $r$-band filter is narrow such that the PSF is not expected to significantly vary across the band.} we conclude that our point-source sample\footnote{Point-source samples can be accessed through the KiDS-DR4 full multi-band catalogue using the column {\sc SG_FLAG}.  Users of the KiDS-1000 gold-shear catalogue need not apply corrections to excise point-sources as they are efficiently removed by the {\it lens}fit weights.} is sufficiently pure from contamination of extended sources to permit accurate PSF modelling.

	\subsection{PSF modelling: Spatial variation}
	\label{sec:psfncoeff}	
	
\begin{figure}
\begin{center}
\includegraphics[width=0.5\textwidth]{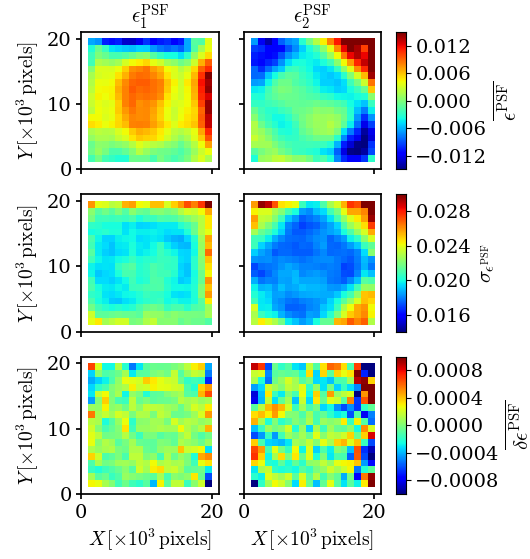}
\caption{Average KiDS-1000 PSF ellipticity $\epsilon^{\rm PSF}$ (upper panels), the associated standard deviation (middle panels), and the residual PSF ellipticity $\delta \epsilon^{\rm PSF}$ (lower panels) on the OmegaCAM focal plane, for the first (left panels) and second (right panels) components of the ellipticity. We note that colour-scale changes between rows.}\label{fig:PSFellip_OnTheCCD}
\end{center}
\end{figure}

	Following 
	\citet{miller/etal:2013} and \citet{kuijken/etal:2015},
	our PSF model
	is defined on a $32 \times 32$ grid of pixels with resolution equal to that of the CCDs (0.213 arcsec per side). The amplitude of each pixel is fit with 
	a two-dimensional polynomial of order $n$, where the coefficients up to order $n_{\rm c}$ are given the freedom to vary between each of the $N_{\rm D}=32$ CCD detectors in OmegaCAM.  This allows for flexible spatial variation (including discontinuities) in the PSF. In each $32\times 32$ pixel grid the amplitudes are  normalised to sum to unity. The total number of model coefficients per pixel is given by \citep{kuijken/etal:2015},
	\begin{equation}
	N_{\rm coeff} = \frac{1}{2} \left[ (n+1)(n+2) + (N_{\rm D}-1)(n_{\rm c} +1)(n_{\rm c}+2) \right] \,.
	\label{eqn:PSFModel_KiDS}
	\end{equation}
	The flux and centroid of each star are also allowed to vary in the fitting, with a sinc function interpolation employed to align the PSF model grid with the star position.
	The total number of coefficients is large, but is sufficiently well constrained by the number of data points, equal to the number of pixels times the number of identified stars in each exposure. Figure~\ref{fig:PSFellip_OnTheCCD} presents the average PSF pattern $\epsilon_{\rm PSF}$, the variance in the PSF ellipticity as measured between the 1006 tiles that comprise KiDS-1000, and the average PSF residual, the difference between the measured PSF ellipticity and the model at the location of the stars, 
	\be
	\delta \epsilon_{\rm PSF} = \epsilon^{\rm PSF}_{\rm true} - \epsilon^{\rm PSF}_{\rm model} \,.
	\ee
These diagnostics are shown for both components of the PSF ellipticity, $\epsilon_1$ (left), and $\epsilon_2$ (right).	  The mean PSF ellipticity is at the percent level, with a standard deviation at the few percent level.  The strongest PSF distortion is seen at the edges of the field of view.   These edge distortions are somewhat mitigated, however, as the KiDS dither strategy is such that the outer $\sim 5 \%$ of all field edges are excised, 
with the area appearing in a more central overlap region 
in an adjacent KiDS-1000 pointing.   

One route to test the reliability of the PSF model is to separate the stellar sample into a training and validation set, where the PSF model is created from the training sample, and the residual PSF ellipticities are determined for the validation set \citep[see for example][]{jarvis/etal:2016}.   We do not adopt this approach, however, as we found that it serves to significantly degrade our PSF model in low-stellar density regions; constraining the high number of coefficients in our model requires the maximum number of stars for the fit.    In our analysis the training and validation sample are therefore the same. 
This choice restricts the direct detection of `over-fitting' in the PSF model which, if present, would imprint PSF ellipticity and size residuals on the inferred galaxy shapes. Our galaxy catalogue tests for the amplitude of such a systematic (Sect.~\ref{sec:H06sysmod} and Sect.~\ref{sec:bmodes}), however, indirectly demonstrate that if any PSF model over-fitting is present, the impact is within our requirements.

	For KiDS-1000 we retain the per-chip coefficient of $n_{\rm c}=1$, as in previous analyses.   
	However, we found that we were able to increase 
	the two-dimensional polynomial 
	order across the field of view from $n=3$
	\citep[used in][]{kuijken/etal:2015, hildebrandt/etal:2017}	
	to $n=4$.  With the improved Gaia-selected point-source sample, the additional coefficients for this higher-order model were found to be well constrained.  This enhancement resulted in a reduction, by a factor of roughly two, in the amplitude of the  residual PSF ellipticity component, $\delta \epsilon_2^{\rm PSF}$, in the upper-right corner of the field of view.   This corner residual is now found at a lower significance, as seen in Fig.~\ref{fig:PSFellip_OnTheCCD}.  Further discussion of the PSF model optimisation is presented in Sect.~\ref{sec:phmodelinkids}. 	  

\subsection{Quantifying the impact of PSF residuals with the Paulin-Henriksson et al. systematics model}
\label{sec:PHmain}
	\citet[hereafter `PH08']{paulin-henriksson/etal:2008} quantify the impact of residual PSF ellipticity and size on cosmic shear estimates for a shear estimator, $e_{\rm obs}$, that is given by
	\be
	e_{\rm obs} = \frac{e_{\rm raw} T_{\rm raw} -  e_{\rm PSF} T_{\rm PSF} }{T_{\rm raw} - T_{\rm PSF} } \, .
	\label{eqn:phmodel}
	\ee
Here $e$ is the `polarisation',  measured from the second moments of the surface brightness profile via
\be
e_{\rm PSF} = \frac{Q_{11} - Q_{22} + 2\rm{i}Q_{12}}{Q_{11}+Q_{22}} \,, 
\ee
where the quadrupole moment, $Q_{ij}$, is given in Eq.~(\ref{eqn:Moment_Brightness_Profile}), and the object size\footnote{In the literature, $T$ is also referred to as $R^2$, the radius-squared.} is given by $T = Q_{11}+Q_{22}$.  
Measurements are made of the PSF-convolved galaxy light distribution, $e_{\rm raw}$ and $T_{\rm raw}$, with the PSF polarisation, $e_{\rm PSF}$, and size, $T_{\rm PSF}$,  at the location of the galaxy, inferred from the measurements around point-sources in the field.    For the shear estimator in Eq.~(\ref{eqn:phmodel}) to hold, the quadrupole moment weight function in Eq.~(\ref{eqn:Moment_Brightness_Profile}),
$w[I(\boldsymbol{\theta})]=1 \forall \boldsymbol{\theta}$.  In the case of realistic noisy imaging data, however, unweighted quadrupole moments formally lead to infinite noise in the shear estimator.   This motivates the use of a Gaussian weight function to isolate each object \citep{kaiser/etal:1995}, and \citet{massey/etal:2013} discuss the additional scaling factors needed to account for the bias introduced by this Gaussian weight.

At this point it is relevant to note that the PH08 choice of shear estimator, $e_{\rm obs}$,  is not fully representative of the model-fitting {\it lens}fit shear estimator $\epsilon$.   The relationship between the polarisation shear estimator $e_{\rm obs}$ and shear $\gamma$ is given by  $ \langle e_{\rm obs} \rangle \simeq 2(1 - \sigma_e^2)\, \gamma $, to first order, where $\sigma_e^2$ is the per-component variance of the unlensed, noise-free, intrinsic polarisation estimates \citep{schneider/seitz:1995}.    This can be contrasted with the ellipticity shear estimator $\epsilon$, which is related to the shear $\gamma$ as $ \langle\epsilon \rangle = \gamma$, to first order \citep{seitz/schneider:1997}.   This model nevertheless allows us to form a framework to provide an indicative estimate of the impact of PSF modelling errors in our analysis. 

Errors in the modelled PSF size and ellipticity can be assessed through a first-order Taylor series expansion of 
Eq.~(\ref{eqn:phmodel}) (see appendix A of PH08) where
	\be
	e_{\rm obs} \simeq e_{\rm obs}^{\rm perfect} + (e_{\rm obs}^{\rm perfect} -e_{\rm PSF}) \frac{\delta T_{\rm PSF}}{T_{\rm gal}} -\frac{T_{\rm PSF}}{T_{\rm gal}} \delta e_{\rm PSF} \, .
        \label{eqn:phsysmodel}
	\ee
        Here $e_{\rm obs} ^{\rm perfect}$ is the perfect systematics-free shear estimator, $T_{\rm gal}$ is the true size of the galaxy, meaning the measured size in the absence of a PSF convolution.   Errors in the PSF model are quantified through  $\delta T_{\rm PSF}$ and  $\delta e_{\rm PSF}$, the offset between the true PSF and the model PSF  at the location of the galaxy, for example $\delta T_{\rm PSF} := T_{\rm PSF} - T_{\rm model}$.

Cosmic shear is traditionally detected using the two-point shear correlation function estimated from the $\epsilon$-shear estimator as		
	\be
	\hat{\xi}_{\pm}(\theta) = \frac{\Sigma_{i,j} w_i w_j (\epsilon^{\rm obs,i}_t \epsilon^{\rm obs,j}_t \pm \epsilon^{\rm obs,i}_\times \epsilon^{\rm obs,j}_\times)  \Delta_{ij}(\theta)}{\Sigma_{i,j} w_i w_j \Delta_{ij}(\theta)} \, ,
        \label{eqn:twoptest}
	\ee
where the $w$-weighted sum over the tangential, $\epsilon_t$, and cross, $\epsilon_\times$, components of the observed ellipticities is taken over all galaxies $i,j$.  The angular binning function $\Delta_{ij}(\theta) = 1$ when the angular separation between galaxies $i$ and $j$ lies within the bin centred on $\theta$, and is zero otherwise.   We can use this estimator to construct a two-point shear correlation function\footnote{We 
use short-hand notation where $\langle a b \rangle$ denotes the two-point correlation function estimator $\xi_{\pm}(\theta)$ in Eq.~(\ref{eqn:twoptest}), but with the $\epsilon$-terms labelled as sample $i$, replaced with the complex quantity $a$.  The $\epsilon$-terms terms in sample $j$ are then replaced with the complex quantity $b$. For scalar quantities, namely size measurements, the notation $\overline{T}$ denotes the {\it lens}fit weighted value of the scalar quantity $T$, averaged over the full survey.}   
with the systematics model in 
Eq.~(\ref{eqn:phsysmodel}) as\footnote{Here we do not include any null terms which correlate statistically independent quantities, for example $e_{\rm PSF}$ and $e_{\rm obs}^{\rm perfect}$.  We also remove the vanishingly small second-order term $\left[ \overline{\delta T_{\rm PSF}} / T_{\rm gal} \right]^2 \left<e_{\rm obs}^{\rm perfect} e_{\rm obs}^{\rm perfect} \right>$.   We choose to ignore any potential correlations between galaxy shape, $e_{\rm obs}^{\rm perfect}$, and size, $T_{\rm gal}$, which would lead to a position-dependent multiplicative bias.  \citet{kitching/etal:2019} distinguishes between spatially varying and constant sources of bias which impact the first term in Eq.~(\ref{eqn:pherror}). Provided that spatially varying biases in the model PSF size are small, however, they conclude that the average bias, as given in Eq.~(\ref{eqn:pherror}), is sufficient to model the multiplicative errors for the two-point shear correlation function.},

        \eqa{ \label{eqn:pherror}
          \langle e_{\rm obs} e_{\rm obs} \rangle \simeq& \left( 1 + 2 \left[{\frac{\overline{\delta T_{\rm PSF}}}{T_{\rm gal}}}\right] \right) \left< e_{\rm obs}^{\rm perfect} e_{\rm obs}^{\rm perfect} \right> \\ \nn
         + & \, \left[ \,\overline{\frac{1}{T_{\rm gal}}}\,\right] ^2 \left< (e_{\rm PSF} \, \delta T_{\rm PSF}) \,  (e_{\rm PSF} \, \delta T_{\rm PSF}) \right>\\ \nn
         +2 & \, \left[ \,\overline{\frac{1}{T_{\rm gal}}}\,\right] ^2 \left< (e_{\rm PSF} \, \delta T_{\rm PSF}) \,  (\delta e_{\rm PSF} \, T_{\rm PSF}) \right>\\ \nn
          + & \, \left[ \,\overline{\frac{1}{T_{\rm gal}}}\,\right] ^2 \left< (\delta e_{\rm PSF} \, T_{\rm PSF}) \,  (\delta e_{\rm PSF} \,T_{\rm PSF}) \right> \, .
        }          
        We note that Eq.~(\ref{eqn:pherror}) differs from similar derivations in 
\citet{massey/etal:2013}, \citet{melchior/etal:2015} and \citet{jarvis/etal:2016}, 
as we choose to keep all terms that may couple within the correlation function.   Specifically we include the possibility where errors in the PSF polarisation, $\delta e_{\rm PSF}$, are correlated with the PSF size, $T_{\rm PSF}$.   Furthermore, \citet{jarvis/etal:2016} choose to link the PH08 systematics model in Eq.~(\ref{eqn:phsysmodel}) with a first-order systematics model (see Sect.~\ref{sec:H06sysmod}) by connecting the $(\delta T_{\rm PSF} / T_{\rm gal}) e_{\rm PSF}$ term with a fractional residual PSF 
        term $\alpha e^{\rm PSF}$ measured directly from the data.   
        We discuss this further in Sect.~\ref{sec:H06sysmod}.  We recognise the third and fourth terms in Eq.~(\ref{eqn:pherror}), as the \citet{rowe:2010} statistics, scaled by PSF and galaxy size ratios.   These terms join the second term in acting as an additive shear bias, in contrast to the first systematic term in Eq.~(\ref{eqn:pherror}), which acts as a multiplicative shear bias.

        \subsubsection{Constraints on the Paulin-Henriksson et al. model}
        \label{sec:phmodelinkids}
        
        We measure each term in Eq.~(\ref{eqn:pherror}) directly from the data, with the exception of the perfect systematics-free shear estimator term, $\langle e_{\rm obs}^{\rm perfect} e_{\rm obs}^{\rm perfect} \rangle$. This term is given by the theoretical expectation for $\xi_+(\theta)$ for the KiDS-1000 redshift distributions from \citet{hildebrandt/etal:2020b} and our fiducial set of cosmological parameters.

        We calculate the PSF polarisation, $e_{\rm PSF} $, and 
        size,
        $T_{\rm PSF}$,  for each object in our stellar sample using a weight function, $w[I(\boldsymbol{\theta})]$ in Eq.~(\ref{eqn:Moment_Brightness_Profile}), given by an iteratively centred Gaussian of width 0.5 arcseconds.  This weight function is necessary to minimise the impact of noise in the wings of the PSF.    To be consistent, we apply the same weight function in the model PSF measurements, 
        $e_{\rm model} $ and $T_{\rm model}$, 
        even though the PSF model is noise-free.  For a circular Gaussian PSF, the weighted and unweighted polarisation measurements are equal in the absence of noise.   For the low-ellipticity seeing-dominated OmegaCAM PSFs, we are relatively close to this regime.
The weighted size estimate is, however, artificially reduced in comparison to the unweighted size estimate with the size of the reduction dependent on the relative size of the PSF to the width of the weight function \citep{duncan/etal:2016}.   \citet{massey/etal:2013} estimate that for small galaxies, weighted size estimates are underestimated by a factor of approximately two, and we adopt this factor to roughly correct our $T_{\rm PSF}$ size estimates.  We also divide each PSF polarisation term, $e_{\rm PSF}$, by a factor of two
to account for the factor of approximately two in the polarisation-shear relation for this shear estimator \citep{schneider/seitz:1995}.  

We estimate the average unconvolved galaxy size, 
$\overline{1/T_{\rm gal}}$, by taking the {\it lens}fit-weighted average of $T_{\rm gal} =6r_{\rm s}^2$ where $r_{\rm s}$ is the exponential disk scalelength of each galaxy as determined from the best-fit galaxy model.    The factor of six results from the requirement for consistent size definitions between the PSF and galaxy, and is calculated analytically from Eq.~(\ref{eqn:Moment_Brightness_Profile}), with the 2D surface brightness profile $I(\theta) \propto \exp\left( -\theta/r_{\rm s} \right)$, following an exponential profile of scalelength $r_s$.    We argue that this approach is an improvement over the alternative of fixing $\overline{T_{\rm PSF}/T_{\rm gal}} = 1$ \citep{jarvis/etal:2016,mandelbaum/etal:2018}, which inappropriate for 
a {\it lens}fit weighted approach,
where small galaxies 
are downweighted in the analysis. 

We compute $\delta \xi_{\pm} = \langle e_{\rm obs} e_{\rm obs} \rangle -\left< e_{\rm obs}^{\rm perfect} e_{\rm obs}^{\rm perfect} \right>$, using the same short-hand notation from Eq.~(\ref{eqn:pherror}).   The correlations are measured using Eq.~(\ref{eqn:twoptest}), with a unit weight for all stars.   The OmegaCAM PSF has equal tangential and radial distortions 
\citep[see for example,][]{kuijken/etal:2015}, 
such that we find $\xi_-^{\rm PSF, PSF}(\theta)$ to be consistent with zero for all $\theta$.   We therefore limit our systematics analysis to the $\xi_+(\theta)$ correlation function.

We analyse four different PSF models characterised by the polynomial orders $n$:$n_{\rm c}$ $=$ 3:1, 4:1, 3:2 and 5:1 (see Eq.~\ref{eqn:PSFModel_KiDS}).  We find that although the PSF is accurately modelled in all four cases, as determined by the low-level measurement of $\delta \xi_+(\theta)$, the 3:1 model, used for the previous KiDS data releases \citep{kuijken/etal:2015, hildebrandt/etal:2017}, does not perform as well as the other three cases.  The level of systematics indicated by the PH08 model are very similar for the 4:1, 3:2 and 5:1 models and we therefore adopt the 4:1 model for KiDS-1000, given that it has the least number of coefficients.

\begin{figure*}
\begin{center}
\includegraphics[width=\textwidth]{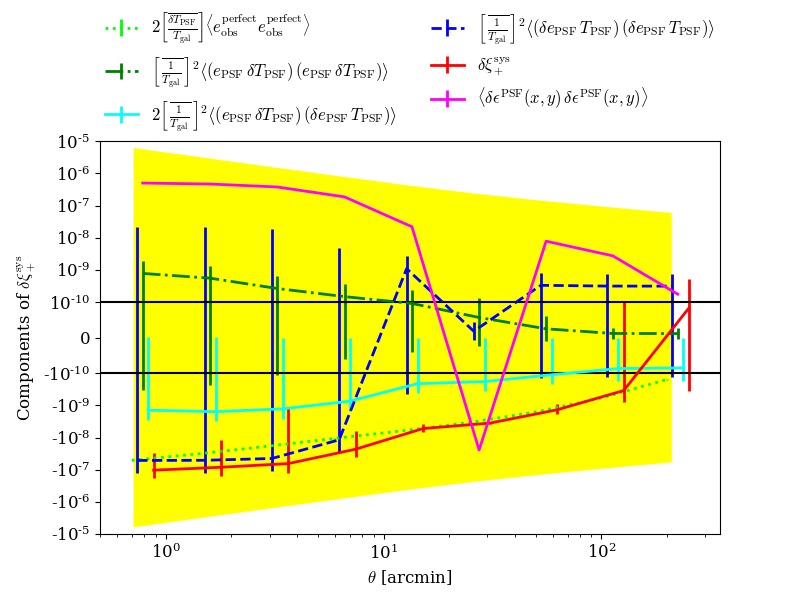} 
\caption{Contributions to the additive systematic, $\delta\xi_+^{\rm sys}(\theta)$, from the PH08 systematics model. The four terms from Eq.~(\ref{eqn:pherror}), shown in varying shades of blue/green (see legend for details), cause $\langle e_{\rm obs} e_{\rm obs} \rangle$ to deviate from $\langle e_{\rm obs}^{\rm perfect} e_{\rm obs}^{\rm perfect} \rangle$. The total systematic $\delta\xi_+^{\rm sys}$ (red), given by the summation of these four terms, can be compared to the yellow band which encloses half the uncertainty on $\xi_+$, assuming a non-tomographic cosmic shear analysis.  As the correlations can be negative, the vertical axis has a symlog scale with black lines indicating the transition from the logarithmic to linear scale. The apparent asymmetry of the error bars (computed via jackknife realisations) is just an artefact of them crossing the log-linear scale boundary.  The PH08 systematics model (red) can be compared to the magenta curve, which shows the expected contribution to the cosmic shear signal from the detector-level bias found in 3/32 OmegaCAM CCDs 
(see Sect.~\ref{sec:flux_additive}).   
This figure presents the analysis of the fifth tomographic bin; similar results are obtained for the other bins. } \label{fig:deltaxip}
\end{center}
\end{figure*}

The measured $\delta \xi_+ (\theta)$, and the 
amplitudes 
of the different contributing terms in Eq.~(\ref{eqn:pherror}), 
are shown in Fig.~\ref{fig:deltaxip}
for the fifth tomographic bin, with similar results found for the other bin combinations.  All the individual contributing terms, and the sum, 
are
within the \cite{mandelbaum/etal:2018} defined requirement limits, shown as a yellow shaded region, and discussed further in Sect.~\ref{sec:requirements}.  We note that Fig.~\ref{fig:deltaxip} uses a symlog scale for the vertical axis as these statistics can be negative; the transition from logarithmic to linear is indicated by the solid horizontal lines. The error bars (too small to be seen in some cases) come from a jackknife resampling, whereby the field is divided into $N_{\rm jk}=49$ segments which are removed one-by-one, with the different terms calculated from the remaining $N_{\rm jk}-1$ segments at each iteration. 

Figure~\ref{fig:deltaxip} shows that the most dominant systematic derives from the first term in Eq.~(\ref{eqn:pherror}) (shown dotted) which is a multiplicative bias, arising from PSF size modelling errors. For tomographic bins 2, 4 and 5, we find $\overline{\delta T_{\rm PSF} /T_{\rm gal}} \sim  (-2 \pm 0.02)\times 10^{-4}$.  This term is consistent with zero, however, for tomographic bins 1 and 3.   
The average residual size modelling error is taken over the full point-source sample, 
and the reported error on the mean 
does not include any errors that arise from the flux-dependent biases discussed in 
Sect.~\ref{sec:flux_multiplicative}.
We remind the reader that the value is also dependent on the size of the weight function used to estimate $T_{\rm PSF}$ in Eq.~(\ref{eqn:Moment_Brightness_Profile}).  We currently only approximately account for the impact of this weight function using the `small-galaxy' correction factors from \citet{massey/etal:2013}.   Nevertheless we note that
in the \citet{asgari/etal:2020b} KiDS-1000 cosmic shear analysis, we marginalise over the uncertainty in the calibration bias correction $m$, per tomographic bin $i$, with $\delta_m^i \sim 0.01 - 0.02$ (see Table~\ref{tab:catstats}).   As the calibration correction to the shear correlation function $\xi_\pm^{ij}(\theta)$, given by $(1+\delta_m^i)(1+\delta_m^j)$, is larger than the measured amplitude of the first term in Eq.~(\ref{eqn:pherror}),  we conclude that the $\delta_m$-marginalisation will mitigate the presence of the multiplicative systematics that we find associated with PSF size modelling errors.

\subsubsection{Accuracy requirements for the PSF model}
\label{sec:requirements}
The procedure for establishing requirements for the PSF modelling, in terms of the additive bias $\delta\xi_+$, is an open question \citep[see for example the discussion in][]{kitching/etal:2019}. \citet{zuntz/etal:2018} note that requirements are specific to individual science cases, but provide a guide that it should be less than 10\% of the weakest tomographic cosmic shear signal.  \cite{mandelbaum/etal:2018} set the requirement that each of the individual terms in Eq.~(\ref{eqn:pherror}) will not exceed $0.5\sigma_{\xi_+}$, where $\sigma_{\xi_+}$ is the standard deviation of $\xi_+$ in each tomographic bin.

Figure~\ref{fig:deltaxip} compares the amplitude of each term in Eq.~(\ref{eqn:pherror}) to $0.5\sigma_{\xi_+}$ (yellow band) where, in contrast to \cite{mandelbaum/etal:2018},  we take $\sigma_{\xi_+}$ to be the error for a non-tomographic analysis.  As we find the PSF errors contaminate each tomographic bin fairly equally, we argue that any requirements based on the measured noise on the correlation function, $\sigma_{\xi_+}$, must use a more stringent requirement set by the noise from a non-tomographic analysis.   We find that the level of systematics in KiDS-1000, as predicted by PH08, meets this requirement.

\cite{troxel/etal:2018} verify that their measured amplitude of $\delta\xi_+(\theta)$ does not significantly bias the DES Year 1 cosmological constraints, through a parameter inference analysis of a biased mock cosmic shear data vector.     We adopt a similar philosophy, but given the computational expense of full MCMC parameter inference analyses we introduce an initial rapid $\chi^2$ analysis of a series of noisy mock data vectors to first flag problematic systematic signals.  Any systematics that raise a flag are referred to the 
computation-intensive
MCMC analysis in order to quantify the resulting bias on the cosmological parameters. 

For our rapid $\chi^2$ test we define the following $\chi^2$-statistics
\begin{equation}
\begin{split}
{\chi^2_{\rm{perfect}}}^j = & \, \bm{\eta}_j^T \, \bm{C}^{-1} \, \bm{\eta}_j  \\
{\chi^2_{\rm{sys}}}^j = &  \, (\bm{\eta}_j + \bm{\delta\xi})^T \, \bm{C}^{-1} \, (\bm{\eta}_j + \bm{\delta\xi})  \\
{\chi^2_{\rm{high/low}}}^j = & \, (\bm{\eta}_j + \bm{\xi}_{\rm{high/low}} - \bm{\xi})^T \, \bm{C}^{-1} \, (\bm{\eta}_j + \bm{\xi}_{\rm{high/low}} - \bm{\xi}) . \\
\end{split}
\label{eqn:chisqtest}
\end{equation}
Here $\bm{\xi}$ is 
the tomographic cosmic shear data vector for the fiducial cosmology, $\bm{\eta}_j$ is the $j$'th realisation ($j \in [0, 5000]$) of the noise on $\bm{\xi}$, sampled from the full KiDS-1000 tomographic cosmic shear covariance matrix $\bm{C}$, described in \citet{joachimi/etal:2020}, $\bm{\delta\xi}$ is the expected systematic bias vector from Eq.~(\ref{eqn:pherror}), replicated for each tomographic bin combination, and finally $\bm{\xi}_{\rm high/low}$ is the tomographic cosmic shear data vector with $S_8$ increased/decreased relative to the fiducial cosmology by a variable factor\footnote{$A$ was chosen to span a reasonable range with $A = 0.1, 0.15, 0.2, 0.3, 0.4$.} of $\pm A\sigma_{S_8}$ of the KiDS-1000 $S_8$ constraint in \citet{asgari/etal:2020b}, where $\sigma_{S_8} \simeq 0.02$.   The $\chi^2_{\rm{perfect}}$ values are those that would be measured for KiDS-1000 in the case of perfect shear measurement across a series of random noise realisations.  The $\chi^2_{\rm{sys}}$ values determine the $\chi^2$ offset introduced when the systematic PSF model bias is included in the measurements, for the same series of noise realisations.    This offset quantifies the reduction in the goodness-of-fit of the perfect cosmological model to the observed signal which includes systematics.   

An initial estimate for the impact of the systematic signal on the inferred cosmological parameters is determined by comparing the offset, $ \Delta \chi^2_{\rm sys} = \chi^2_{\rm{sys}} - \chi^2_{\rm{perfect}}$, to the $\chi^2$ offset introduced when changing the underlying $S_8$ cosmology by $\pm A\sigma_{S_8}$, through a series of different $\chi^2_{\rm{high/low}}$ values. As such, we are assuming that the bias in the goodness-of-fit caused by the systematic, mimics a change in the $S_8$ 
parameter. In reality, however, a given PSF systematic could induce changes in the best-fit values of multiple cosmological and nuisance parameters, or could alter the $\chi^2$ without introducing any bias 
in
the best-fit parameters \citep{amara/etal:2008}.  Our $\chi^2$ test therefore only serves as a benchmark for the impact of a PSF systematic on the inference of the $S_8$ parameter.  
If the systematic is found to induce significant changes in the goodness-of-fit, the systematic can then move up to the next stage of testing using a full MCMC analysis.

Specifically, we calculate the mean of each  $\chi^2$ distribution, and find the lowest amplitude $A$, where the shift between the `perfect' and `sys' hypotheses, $\Delta \chi_{\rm sys}^2$, is smaller than the shifts induced between the perfect and `high' or `low' hypotheses, $\Delta \chi_{\rm high,low}^2$. As the values of $\Delta \chi_{\rm sys,high,low}^2$ vary slightly with the shot noise $\bm{\eta}$, we measure the average shifts over 20 iterations of the $\chi^2$ distributions, each consisting of 5000 noise realisations.
For the systematic bias vector given in Eq.~(\ref{eqn:pherror}), we 
find $\overline{\Delta \chi_{\rm sys}^2} = 0.001 \pm 0.001$ which is smaller than the shifts induced between the perfect and high or low hypotheses with $A=0.1$, where $\overline{\Delta \chi_{\rm high}^2} =0.016 \pm 0.004$ 
and
$\overline{\Delta \chi_{\rm low}^2} =0.022 \pm 0.004$. We therefore conclude that the low-level imperfections in our PSF modelling, seen in Fig.~\ref{fig:deltaxip}, induce a change in the goodness of fit that 
is significantly less than the change induced if the underlying $S_8$ cosmology changes by $0.1 \sigma_{S_8} = 0.002$.   

\citet{joachimi/etal:2020} determine a requirement for systematics to induce a $<0.1\sigma$ change on $S_8$.  This limit corresponds to the typical variance between the values of $S_8$ recovered from a series of converged MCMC parameter inference chains that analyse the same mock KiDS-1000 data vector, but with different random seeds.  Based on the results of our rapid $\chi^2$ analysis, we find that there is no necessity to run an expensive full MCMC analysis to accurately quantify the bias incurred as a result of the presence of the additive bias $\delta\xi_+$ shown in Fig.~\ref{fig:deltaxip}.  At the estimated level of $<0.1\sigma$ differences, any small offsets in the MCMC results could simply be attributed to noise in the parameter estimation.   We therefore conclude that the accuracy of the KiDS-1000 PSF model, as quantified with the PH08 method, is well within our requirements for KiDS-1000.

\subsection{Detector-level effects}
\label{sec:detectorlevel}
The discussion thus far has assumed that the PSF is the only source of instrumental bias, such that, in the absence of noise, 
Eq.~(\ref{eqn:phmodel}) provides an unbiased estimate of the galaxy shape.   Imperfections in the detector are not captured by this equation, however, and they can also introduce biases in the measured shapes of galaxies \citep{massey/etal:2013}.  

One of the best-known examples of a detector-level distortion is `charge transfer inefficiency' (CTI), which is particularly relevant for space-based lensing studies where the background is low \citep[see for example][]{miralles/etal:2005, rhodes/etal:2007}. In this case not all the charge in a pixel is transferred at each readout cycle, and the trapped charge is released some time later, with the release probability determined by the type of defect in the silicon lattice. Traps with release times similar to the clocking time cause a trail that increases with 
a given object's distance
 to the readout register \citep[see for example][]{massey/etal:2010}.  Although prominent in space-based observations, it is a common feature of all CCD detectors.  

The `brighter-fatter effect' (BFE) introduces another distortion.  Here the build-up of charge in a given pixel acts to inhibit said pixel's capture of further incident photons, such that they are captured by the surrounding pixels. This results in a broadened PSF for brighter objects \citep{antilogus/etal:2014}.   The flux dependence of the effect is typically different for the parallel and serial readout directions, modifying both the PSF size and ellipticity as a function of the pixel count value.  

Lesser-known effects include `pixel bounce', which is similar to CTI, and could be caused by dielectric absorption in the read-out electronics \citep{toyozumi/etal:2005}.  Here, capacitors in the circuit do not reset to the reference bias voltage sufficiently quickly. If the voltage is too low, the recorded pixel value is biased high. The excess signal depends on the counts in the previous pixel inducing a distortion along the readout direction.  Unlike CTI, however, the bias in the object shape does not depend on the distance to the readout register.  The trail is also shorter.   
Additionally, there is the 
so-called `binary offset effect' \citep{boone/etal:2018} which results in a shift of charge, by up to three pixels, as a result of the digitisation of the CCD output voltage.  Hoekstra et al. (in prep.) have detected this effect in OmegaCAM data, but conclude that it is irrelevant given the sky background levels in the KiDS $r$-band data. 

These detector-level distortions are all dependent on the flux of the object.  Any model for the PSF derived from measurements of bright stars may therefore be inappropriate for faint galaxies.   A biased PSF correction then leads to biased galaxy shape measurements \citep{melchior/etal:2015}.

Hoekstra et al., (in prep.) present a detailed study of pixel correlations in flat-field exposures from OmegaCAM, detecting increased noise-covariance at bright fluxes, a clear signature of BFE.  Given the thinned OmegaCAM CCDs, however, the fluxes where the effect becomes significant are high, and the impact on the shape and ellipticity of the PSF itself was found to be very small for stars with magnitudes $r>18$.   This bright limit was therefore adopted in our PSF modelling.  In the same analysis, Hoekstra et al. (in prep.) study CTI in the OmegaCAM serial readout direction, detecting a low-level signal that does not vary significantly between detectors.   They conclude, however, that the CTI distortion is at level that does not affect the shape measurements.

\subsubsection{Quantifying PSF flux dependent additive bias}
\label{sec:flux_additive}
\begin{figure}
\includegraphics[width=0.5\textwidth]{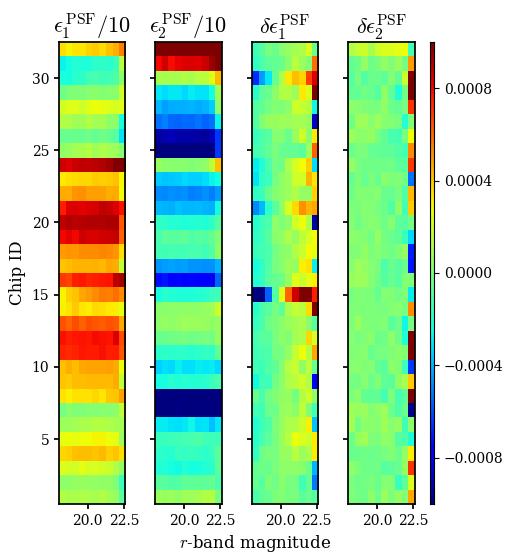} 
\caption{Average KiDS-1000 PSF ellipticity $\epsilon^{\rm PSF}$ (left panels, 
divided by a factor of ten)
 and residual PSF ellipticity $\delta \epsilon ^{\rm PSF}$ (right panels), indicated by the colour bar, as a function of stellar $r$-band magnitude and CCD chip ID.  A flux dependence of the PSF residual $\delta \epsilon_1 ^{\rm PSF}$ is seen in CCD chip IDs 15, 21, and 30, indicating the 
presence of strong detector-level systematics in these three CCDs.} 
\label{fig:chip_mag_residual}
\end{figure}

We investigate detector-level distortions  in Fig.~\ref{fig:chip_mag_residual}, 
which shows the average residual PSF ellipticity, $\delta \epsilon^{\rm PSF}$, as a function of stellar $r$-band magnitude for the 32 OmegaCAM CCD chips for our star sample with $18<r<22.5$.  For the $\delta \epsilon_2^{\rm PSF}$ component, we find very low levels of flux dependence for all of the CCD chips 
ranging from $\langle \delta \epsilon_2^{\rm PSF} \rangle_{(r=18)} = (2 \pm 1) \, \times 10^{-5}$ to $\langle \delta \epsilon_2^{\rm PSF}\rangle_{(r=22)}= (-4 \pm 2) \,\times 10^{-5}$. For the majority of chips, the flux dependence is also low for the $\delta \epsilon_1^{\rm PSF}$ component which ranges from $\langle \delta \epsilon_1^{\rm PSF}\rangle_{(r=18)}= (-2.3 \pm 0.4) \,\times 10^{-4}$ to $\langle\delta \epsilon_1^{\rm PSF}\rangle_{(r=22)}= (2.2 \pm 0.4)\, \times 10^{-4}$.
Three of the OmegaCAM CCD chips in Fig.~\ref{fig:chip_mag_residual} do, however, exhibit significant flux dependence in the $\delta \epsilon_1^{\rm PSF}$ component of the PSF ellipticity, chips with CCD IDs\footnote{The {\sc THELI} CCD naming convention IDs 15, 21, and 30 correspond to the ESO CCD IDs 74, 84, and 91, where the conversion between the two naming schemes is given by,
\begin{equation*}
{\rm ID}_{\rm ESO} = 16\left[ \left( {\rm ID}_{\rm THELI}-1 \right)//8 \right] +73 - {\rm ID}_{\rm THELI} \,,
\end{equation*}
with $//$ designating integer division.} 15, 21, and 30.   

\begin{figure}
\begin{center}
\includegraphics[width=0.5\textwidth]{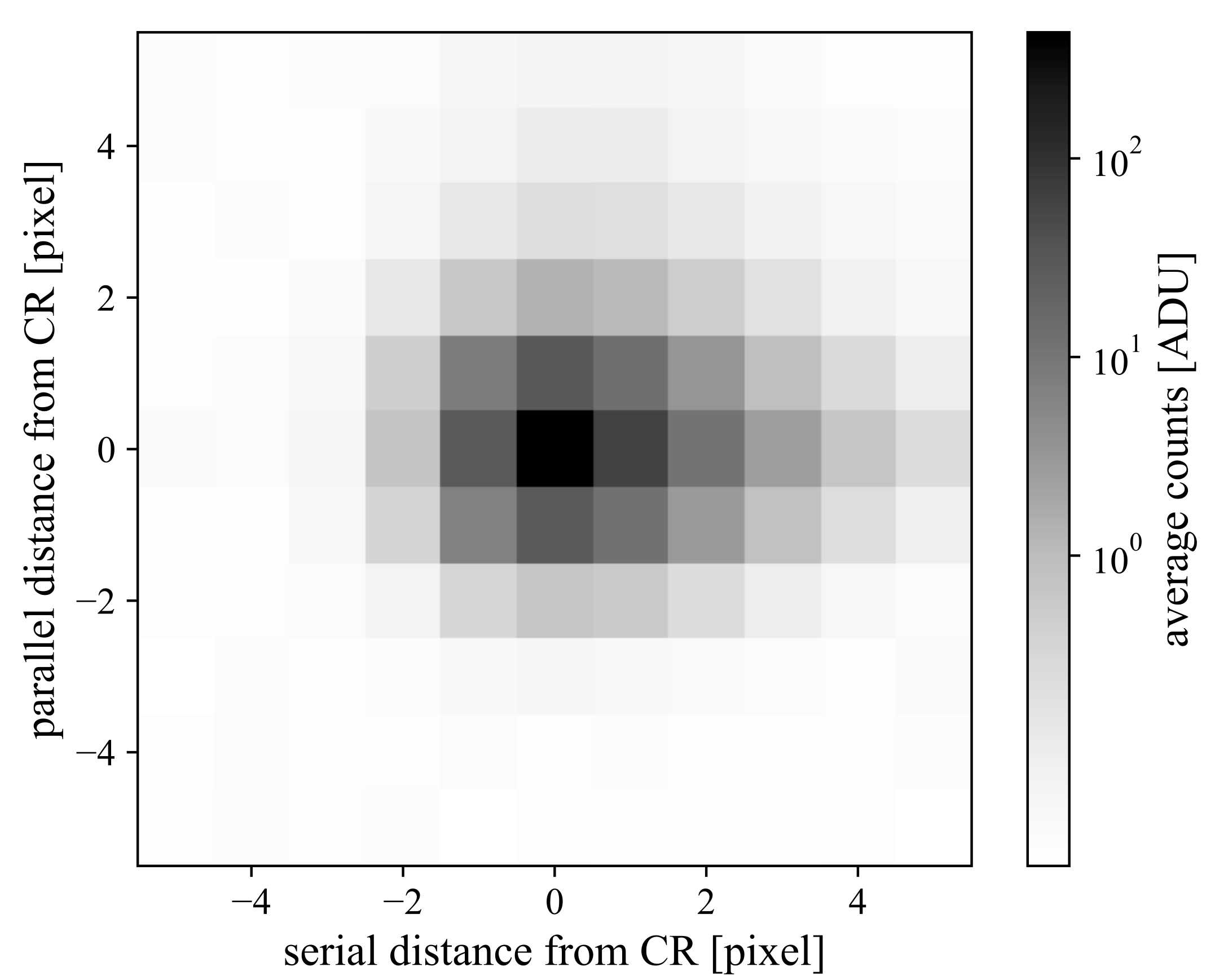} 
\caption{Counts per pixel for a stacked image centred on cosmic rays detected in dark frames from OmegaCAM {\sc THELI} CCD ID 15 (also referred to as ESO CCD ID 74).  A distortion can be seen along the serial direction to the read-out amplifier, which we interpret as primarily arising from pixel bounce.   We note that this effect is found to be significantly lower in all other OmegaCAM detectors.} \label{fig:cr_stack}
\end{center}
\end{figure}

To explore the flux dependence of the PSF further, we analysed cosmic rays in all the OmegaCAM dark frames, which contain no other objects.  Figure~\ref{fig:cr_stack} shows the stack of all cosmic rays with counts between 250 and 800 for the most offending detector, CCD ID 15.  We find trailing in the serial direction, which is flipped in the upper half of OmegaCAM relative to the lower half, supporting a hypothesis that this effect is caused by CTI and/or pixel bounce in the readout register.   By inspecting the dependence on the distance to the readout register, we find that the CTI distortion in this CCD is an order of magnitude smaller than the dominant source of the distortion which we therefore infer arises from pixel bounce.

We model the systematic error introduced by the flux dependence of the PSF seen in Fig.~\ref{fig:chip_mag_residual} following \citet{hildebrandt/etal:2020}, fitting a linear relation to the per-chip $\delta \epsilon^{\rm PSF}$ residual ellipticities as a function of $r$-band magnitude.   We estimate a field-of-view position dependent $\delta \epsilon^{\rm PSF} (x,y)$ model for the typical KiDS galaxy by extrapolating the linear magnitude relationship to $r = 24$.   We mimic the dithering and stacking of exposures, by combining five dithered $\delta \epsilon^{\rm PSF} (x,y)$ maps \citep[see figure 2 in][]{hildebrandt/etal:2020}.   Residual PSF ellipticity contributes to the observed cosmic shear signal with an amplitude $\delta\xi_\pm \approx \langle \delta \epsilon^{\rm PSF} \delta \epsilon^{\rm PSF} \rangle$ 
\citep[see for example the discussion in PH08; ][]{zuntz/etal:2018}.   
We find that |$\delta\xi_+|< 5.1 \times 10^{-7}$, with the angular dependence of the function shown in Fig.~\ref{fig:deltaxip} (magenta curve).    

We quantify the impact of the flux dependence of the PSF distortions on our cosmological parameter constraints using the methodology from Sect.~\ref{sec:requirements}.  We find the change in the goodness-of-fit of the fiducial cosmological model, given a pixel-bounce biased data vector, is consistent with the change in the goodness-of-fit when the value of $S_8$ in the cosmological model changes by $0.15\sigma_{S_8}$, (see Eq.~\ref{eqn:chisqtest}).  This result is consistent with \citet{asgari/etal:2019} who only see a significant impact in their mock data analysis if they artificially increase the magnitude of this detector effect by a factor of five.   This difference is just outside our tolerance requirement of systematics inducing less than a $0.1\sigma_{S_8}$ change in $S_8$, however.  As we discuss further in Sect.~\ref{sec:baconres}, we do not find evidence in the data to support the residual PSF ellipticity model analysed here, with the data favouring a significantly lower amplitude.   We remind the reader that our faint galaxy residual PSF ellipticity model is derived from a linear fit to the effect determined from stars with magnitudes ranging from $18<r<22$, extrapolated to $r=24$.  The fact that the faint galaxy data does not support this extrapolated model is an 
indication that the impact of the effect diminishes as the galaxies approach the background noise level, resulting in a non-linear relationship between the residual PSF ellipticity and galaxy magnitude.

We note that the results of the PH08 model analysis in Sect.~\ref{sec:phmodelinkids} do not predict the level of additive bias anticipated from the estimated detector-level systematic.  This is because our PH08 model analysis is incomplete, as it neglects any flux-dependence in the measured quantities.   We therefore caution that future systematic tests with the PH08 model should build in a flux-dependent dimension, evaluating Eq.~(\ref{eqn:pherror}), as a function of magnitude \citep[see the discussion in][]{massey/etal:2013, cropper/etal:2013}.   We recognise, however, that this development is non-trivial as the various quantities measured in Eq.~(\ref{eqn:pherror}) become progressively noisier as we reach the stellar magnitude limit for the sample of $r\gtrsim22$ (see Fig.~\ref{fig:chip_mag_residual}).  To extend beyond this limit towards typical galaxy magnitudes, we have relied on linear extrapolation, the accuracy of which we test in Sect.~\ref{sec:baconres}.  In the future it is likely that we will become reliant on detailed simulations in order to model and quantify the impact of these systematic effects \citep{paykari/etal:2020}.

\subsubsection{Quantifying PSF flux-dependent multiplicative bias}
\label{sec:flux_multiplicative}
Detector-level distortions impact both the ellipticity and size of the PSF as a function of flux.   As we can see from the first term in Eq.~(\ref{eqn:pherror}), errors in the PSF size, $\delta T_{\rm PSF} = T_{\rm PSF}^{\rm data} - T_{\rm PSF}^{\rm model}$, result in a multiplicative bias on the cosmic shear measurement.  There are, however, two challenges in accurately determining $\delta T_{\rm PSF}$ as a function of stellar magnitude.   The first is a form of noise-bias where, as the PSF becomes progressively fainter, the wings of the distribution dip below the background.  In this case, the $T$-size estimate is unable to distinguish between a narrow high 
surface brightness 
PSF, or an extended lower surface brightness PSF, as the part of the PSF that we observe above the noise threshold appears to be the same size \citep{duncan/etal:2016}.   The second involves the choice of weight function in Eq.~(\ref{eqn:Moment_Brightness_Profile}), which introduces a bias in the recovered 
object's
 size.  This bias depends on the relative size of the object to the weight function, hampering efforts to detect size variation as a function of flux when using a fixed weight size.   Given the difference between $T_{\rm PSF}$
 values
  measured at different fluxes, compared to the chosen weight radius, however, the impact of the weight function bias is expected to be weak.   These challenges currently preclude an accurate quantification of the multiplicative bias that we incur from the weak flux dependence of the PSF seen in Fig.~\ref{fig:chip_mag_residual}.   We can however make a rough calculation based on the data to hand, which is expected to overestimate the amplitude of this effect\footnote{The BFE results in fainter stars that are narrower than bright stars.   The $T$-size estimate underestimates the size of an object as its flux decreases, which mimics BFE.}.

Extrapolating measurements of $\delta T_{\rm PSF}/T_{\rm PSF}$ as a function of stellar $r$-band to $r=24$, we estimate that $-0.005 < \langle \delta T_{\rm PSF}/T_{\rm PSF} \rangle_{(r=24)} < 0$.   At the faint end of the galaxy population, the {\it lens}fit weighted  $\langle T_{\rm PSF}/T_{\rm gal} \rangle_{(r=24)} \sim 1$, where we include the `small-galaxy’ correction factor from \citet{massey/etal:2013} to account for the weight bias in $T_{\rm PSF}$.  Combining these estimates we can conclude that the impact of flux dependent multiplicative bias on the two-point shear correlation function $\xi_\pm$, as quantified through the first term in Eq.~(\ref{eqn:pherror}), is $-0.01<2\langle \delta T_{\rm PSF}/T_{\rm gal} \rangle_{(r=24)} <0$.   The \citet{kannawadi/etal:2019} uncertainty on the calibration correction to the shear correlation function $\xi_\pm^{ij}(\theta)$ is given by $(1+\delta_m^i)(1+\,\delta_m^j)$, where the uncertainties are treated as being 100\% correlated between tomographic bins, with $\delta_m^i$ listed in Table~\ref{tab:catstats}.   As this is a factor of two to four times larger than the measured amplitude of the first term in Eq.~(\ref{eqn:pherror}),  we conclude that the $\delta_m^i$-marginalisation in any cosmic shear analysis will mitigate the presence of the multiplicative systematics that we find associated with 
flux-dependent
 PSF size modelling errors.   

We note that, as in Sect.~\ref{sec:flux_additive}, we have adopted linear extrapolation to model the flux dependence of the residual PSF size $\delta T_{\rm PSF}$.  As faint galaxy data does not support this extrapolated model, in Sect.~\ref{sec:baconres}, the multiplicative bias estimate that we have presented here is very likely to be a worse-case scenario.   It nevertheless
 highlights the 
necessity to develop a new strategy 
for including 
an accurate 
flux-dependent dimension 
in
future systematic tests with the PH08 model.

\subsection{Quantifying the impact of PSF residuals with a first-order systematics model}
\label{sec:H06sysmod}
In this section we directly test for PSF residual ellipticities in the KiDS-1000 shear catalogue using a first-order systematics model applied to the weighted {\it lens}fit galaxy shear estimates.  This is in contrast to the PH08 model analysis from Sect.~\ref{sec:PHmain} which provides an indirect test of the shear catalogue through the PSF model.    Here systematic errors are parameterised using a first-order expansion \citep{heymans/etal:2006} of the form	
\be
\epsilon^{\rm obs} = (1+m)(\epsilon^{\rm int} + \gamma) + \alpha \epsilon^{\rm PSF} + \beta \delta\epsilon^{\rm PSF} + c \,,
\label{eqn:H06sysmod}
\ee
where $\epsilon^{\rm obs}$ is the observed ellipticity, that is the shear estimator,  $m$ is a multiplicative bias, $\epsilon^{\rm int}$ is the intrinsic ellipticity, $\gamma$ is the cosmic shear term that we wish to extract, $\alpha$ and $\beta$ are the fractions of the PSF ellipticity $\epsilon^{\rm PSF}$, and the residual PSF ellipticity $\delta\epsilon^{\rm PSF}$, that remain in the shear estimator, and finally $c$ is an additive term that is uncorrelated with the PSF.  We note that the ellipticity, shear and additive terms in Eq.~(\ref{eqn:H06sysmod}) are written in complex form, for example $\epsilon = \epsilon_1 + {\rm i} \epsilon_2$. The terms $\alpha$, $\beta$ and $m$, however, are typically treated as scalars, scaling both of the ellipticity components equally.  

In the first-order systematics model, a non-zero $\alpha$ can be attributed to an error in the deconvolution of the PSF from the galaxy ellipticities and/or noise-bias. A non-zero $\delta\epsilon^{\rm PSF}$ can be associated with how well the model fits the true effective PSF. \citet{zuntz/etal:2018} argue that in this case, a value of order $\beta \sim -1$ is expected as PSF model errors propagate into an error of the same magnitude, but opposite sign, in the shear estimate.  A non-zero $c$ could be associated with detector level effects such as charge transfer inefficiencies.  

Taking the first-order systematics model from Eq.~(\ref{eqn:H06sysmod}) under the assumption that $m$, $\alpha$ and $c$ are constant across the full survey, we find the two-point shear correlation function estimator $\hat{\xi}_{\pm}$
(see Eq.~\ref{eqn:twoptest}),
 is given by
 \eqa{ \label{eqn:H06error}
\hat{\xi}_{\pm} =& \,   (1+m)^2 \langle \epsilon^{\rm perfect} \epsilon^{\rm perfect} \rangle + \alpha^2 \, \langle \epsilon^{\rm PSF} \epsilon^{\rm PSF} \rangle \\ \nn
 & + 2\alpha\beta \, \langle \epsilon^{\rm PSF} \delta \epsilon^{\rm PSF} \rangle + \beta^2 \, \langle \delta\epsilon^{\rm PSF} \delta\epsilon^{\rm PSF} \rangle + cc_{\pm} \, .
}
Here we follow the short-hand notation from Eq.~(\ref{eqn:pherror}), where, for example, $ \langle \epsilon^{\rm perfect} \epsilon^{\rm perfect} \rangle = {\xi}^{\gamma \gamma}_{\pm}$, the cosmic shear two-point correlation function which is directly related to the non-linear matter power spectrum and its associated cosmological parameters.  We also define $cc_{\pm}$ for the contribution of the scalar $c$-term to  $\hat{\xi}_{\pm}$.  Here $cc_+ = c_1^2 + c_2^2$, and $cc_- = 0$, by definition. 

\citet{bacon/etal:2003} define the following systematics estimator to determine the level of contamination to the two-point shear correlation function estimator $\hat{\xi}_{\pm}$ from any residual PSF ellipticity in the shear estimate
\be
\xi_{\pm}^{\rm sys} =  \langle \epsilon^{\rm obs} \epsilon^{\rm PSF} \rangle^2 /  \langle \epsilon^{\rm PSF} \epsilon^{\rm PSF} \rangle \, ,
\label{eqn:Baconsys}
\ee
where $ \langle \epsilon^{\rm obs} \epsilon^{\rm PSF} \rangle$ is the `star-galaxy' cross-correlation function, measured between the observed and PSF ellipticities.  If the model in Eq.~(\ref{eqn:H06sysmod}) provides a good representation of the systematics in the data, then $\xi_{\pm}^{\rm sys} = \alpha^2   \langle \epsilon^{\rm PSF} \epsilon^{\rm PSF} \rangle$ when $\delta\epsilon^{\rm PSF} \sim 0$.   We note that any significant additive biases, $c$, do not contribute to the $\xi_{\pm}^{\rm sys}$ estimator as, by definition, $c$ is uncorrelated with the PSF.

\subsubsection{Constraints on the parameters of the first-order systematics model}
\label{sec:cterm}
\begin{figure}
\begin{center}
\includegraphics[width=0.5\textwidth]{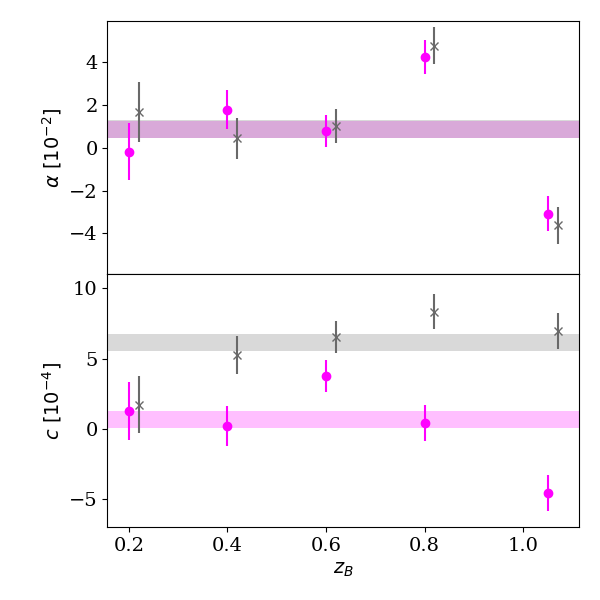}
\caption{Systematics parameters:  the amplitude of the PSF residual fraction $\alpha$, and the additive parameter $c$, from Eq.~(\ref{eqn:H06sysmod}), as a function of tomographic photometric redshift bin, $z_{\rm B}$.   The tomographic measurements can be compared per ellipticity component, $\epsilon_1$ (closed, pink) and $\epsilon_2$ (crosses, grey), and with a non-tomographic measurement (shown as coloured grey/pink bars of width $1\sigma$).}\label{fig:alphac}
\end{center}
\end{figure}

We constrain the amplitude of the PSF leakage term $\alpha$, and the additive parameter $c$, in Fig.~\ref{fig:alphac}, by fitting Eq.~(\ref{eqn:H06sysmod}) to the $w$-weighted KiDS-1000 shear measurements, in the case of position-independent parameters.   In this analysis we note that for the large KiDS area, $\langle \epsilon^{\rm int} + \gamma \rangle \approx 0$.  We also fix $\delta \epsilon^{\rm PSF} = 0$, which is a good approximation
when considering the average PSF modelling error across the full survey (see Fig.~\ref{fig:PSFellip_OnTheCCD}).  We refer the reader to Sect.~\ref{sec:detectorlevel}, however, where we quantify the impact of low-level flux dependent PSF modelling errors in the three out of the 32 CCDs in OmegaCAM which display 
strong detector level effects.    

We find that $\alpha$ is consistent with zero, at the $2\sigma$ level, when considering the full survey (see the shaded region in Fig.~\ref{fig:alphac}), and also the first three tomographic photometric redshift bins with $z_{\rm B}<0.7$. 
For the two highest redshift bins we find $|\alpha| \sim 0.04 \pm 0.01$, with consistent PSF residual fractions when considering the $\epsilon_1$ and $\epsilon_2$ components independently. This validates our model in which $\alpha$ is treated as a scalar, modulating both ellipticity components equally.     

Turning to the additive bias $c = c_1 + {\rm i} c_2$, we find $c_1$ is consistent with zero when considering the full survey (see the pink shaded region in Fig.~\ref{fig:alphac}), but with significant detections in the third and fifth tomographic bins.   We find a significant detection of $c_2$ at the level of $c_2 \sim (6 \pm 1) \times10^{-4}$ across all but the first tomographic bin.   The presence of a significant $c_2$ term is expected from {\it lens}fit analyses of image simulations where \citet{kannawadi/etal:2019} measure $c_2$ in the range of $[5,10] \times 10^{-4}$.    From this image simulation analysis we can conclude that a low-level additive systematic bias, that is uncorrelated with the PSF, is inherent to the software that we have used\footnote{We note that {\it lens}fit remains under development, and the origin of the $c_2$ term is now known to lie in the likelihood sampler.   An updated version of {\it lens}fit will be used for the analysis of the full KiDS survey area (Data Release 5).}.   This is supported by an analysis of the shear catalogue as a function of object position within the field of view, finding no significant variation of $c$ across the camera.

Based on these results, we choose to empirically correct the observed shear estimates such that $\epsilon^{\rm obs}_{\rm corr} = \epsilon^{\rm obs} - \overline{ \epsilon^{\rm obs}}$, where $\overline{ \epsilon^{\rm obs}}$ is the weighted average ellipticity of the relevant tomographic bin.   It is these empirically corrected shear estimates that we use in the cosmic shear analysis of \citet{asgari/etal:2020b}, which includes a nuisance parameter $\delta \overline{\epsilon}^2$ for each tomographic bin to marginalise over our uncertainty in the accuracy of the empirical calibration correction.   The prior for $\delta \overline{\epsilon}^2$ is given by a zero-mean Gaussian of width $\sigma= 7.5 \times 10^{-8}$, corresponding to the largest variance, measured from any tomographic bin, between 300 bootstrap sample measurements of $\langle \epsilon_1^{\rm obs} - \overline{ \epsilon_1^{\rm obs}} \rangle^2 + \langle \epsilon_2^{\rm obs} - \overline{ \epsilon_2^{\rm obs}} \rangle^2$.    We note that this prior is roughly twice the size of the nuisance prior adopted in \citet{hildebrandt/etal:2020}, as it accounts for correlations between the empirical corrections that were previously neglected.

In the first three tomographic bins where $\alpha \sim 0$, the empirical correction that we apply corresponds to the additive $c$-term, that is $\overline{ \epsilon^{\rm obs}} \equiv c$.   In the highest two tomographic bins, however, $\overline{ \epsilon^{\rm obs}} \equiv \alpha \overline{ \epsilon^{\rm PSF}} + c $.  This correction therefore also accounts for the average offset induced from the PSF residual ellipticities which is of similar amplitude to the $c$-terms\footnote{The average PSF ellipticity is $\overline{ \epsilon_1^{\rm PSF}} = 0.005 \pm 0.001$, $\overline{ \epsilon_2^{\rm PSF}} = -0.005 \pm 0.001$ for the KiDS-1000 equatorial field, and $\overline{ \epsilon_1^{\rm PSF}} = 0.003 \pm 0.001$, $\overline{ \epsilon_2^{\rm PSF}} = 0.001\pm 0.001$ for the KiDS-1000 southern field.}.   We note that we choose not to implement an additional empirical PSF-dependent correction for the PSF contamination, as our $\alpha$ measurements are noisy.  Furthermore, if $\alpha$ is not a constant, and is instead correlated with galaxy properties, atmospheric seeing, or image depth, for example, applying an average correction would artificially imprint a PSF-correlation across our full data set that could potentially be more problematic than the low-level average bias that we currently detect. 

\subsubsection{Accuracy requirements and validation of the first-order systematics model}
\label{sec:baconres}

\begin{figure}
\includegraphics[width=0.5\textwidth]{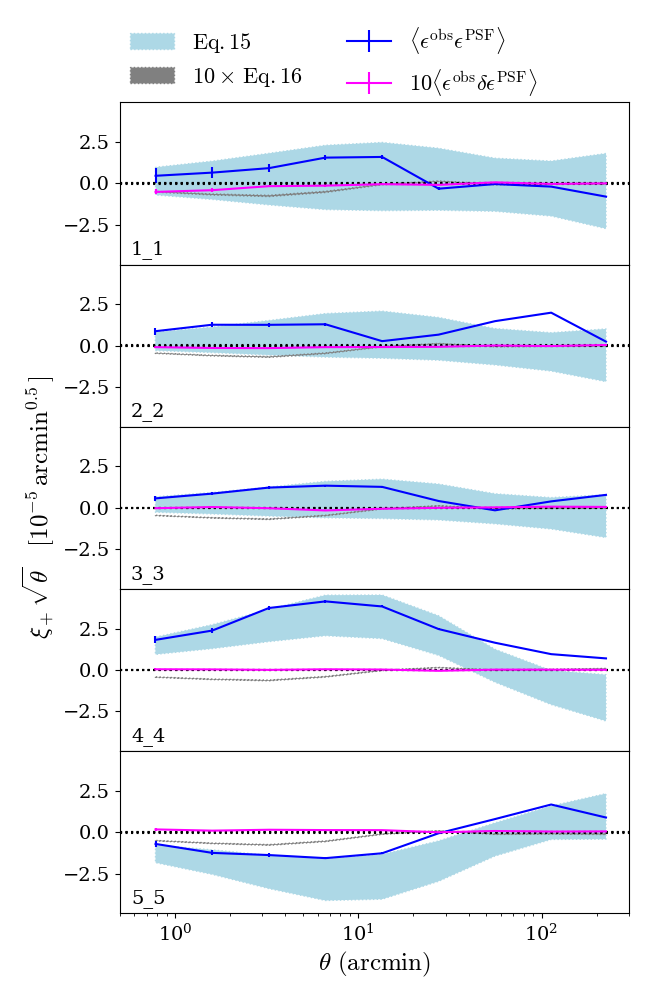} 
\caption{Comparison of two star-galaxy cross-correlation functions with their corresponding linear systematics model predictions (Eqs.~\ref{eqn:gemod} and \ref{eqn:gdemod}) for the five tomographic bins.    The star-galaxy cross correlation function $\langle \epsilon^{\rm obs}_{\rm corr} \, \epsilon^{\rm PSF} \rangle$ (dark blue) is in reasonable agreement with the model (blue bar), where the width of the model reflects the $2\sigma$ uncertainty.  The residual star-galaxy cross correlation function $\langle \epsilon^{\rm obs}_{\rm corr} \, \delta \epsilon^{\rm PSF} \rangle$ (shown pink), and corresponding model (grey), are scaled by a factor of ten in order to display the measurements on the same scale.   We also scale the correlation functions by $\sqrt{\theta}$ to aid visualisation of the large-scale signal.}
\label{fig:stargal}
\end{figure}

In Fig.~\ref{fig:stargal} we compare two measured star-galaxy cross-correlation functions with the amplitudes predicted by the linear systematics model from Eq.~(\ref{eqn:H06sysmod}).    The standard star-galaxy cross-correlation function,  $\langle \epsilon^{\rm obs}_{\rm corr} \, \epsilon^{\rm PSF} \rangle$, is related to the linear systematic model parameters as\footnote{Here we use the same notation as Eq.~(\ref{eqn:H06error}), with $ab_\pm = a_1b_1 \pm a_2b_2$.  We also adopt the notation from Eq.~(\ref{eqn:pherror}), where $\overline{\epsilon_i}$ indicates the {\it lens}fit weighted average value of the scalar quantity $\epsilon_i$, and $\overline{\epsilon} =\overline{\epsilon_1} + {\rm i} \overline{\epsilon_2}$ .  For completeness we remind the reader that $\epsilon^{\rm obs}_{\rm corr} = \epsilon^{\rm obs} - \overline{ \epsilon^{\rm obs}}$.}
\be
\langle \epsilon^{\rm obs}_{\rm corr} \, \epsilon^{\rm PSF} \rangle = \alpha \langle \epsilon^{\rm PSF} \epsilon^{\rm PSF} \rangle + \beta \langle \epsilon^{\rm PSF} \delta\epsilon^{\rm PSF} \rangle - \alpha \overline{ \epsilon^{\rm PSF}}\,\overline{ \epsilon^{\rm PSF}}_\pm \, .
\label{eqn:gemod}
\ee

We can also construct a residual star-galaxy cross correlation function $\langle \epsilon^{\rm obs}_{\rm corr} \, \delta \epsilon^{\rm PSF} \rangle$, where
\be
\langle \epsilon^{\rm obs}_{\rm corr} \, \delta \epsilon^{\rm PSF} \rangle = \alpha \langle \epsilon^{\rm PSF} \delta \epsilon^{\rm PSF} \rangle + \beta \langle \delta \epsilon^{\rm PSF} \delta \epsilon^{\rm PSF} \rangle - \alpha \overline{ \epsilon^{\rm PSF}} \, \overline{ \delta\epsilon^{\rm PSF}}_\pm \, .
\label{eqn:gdemod}
\ee
As with previous sections, we focus on the $\xi_+(\theta)$ term only, as the $\xi_-(\theta)$ term is consistent with zero for the OmegaCAM PSF.

Figure~\ref{fig:stargal} shows reasonable agreement between the measured star-galaxy correlation function (shown dark blue), with the model prediction (blue band, Eq.~\ref{eqn:gemod}) which we calculate by taking $\alpha$ from Fig.~\ref{fig:alphac}, $\beta = -1$, $\delta \epsilon^{\rm PSF}$ from the detector level model in Sect.~\ref{sec:detectorlevel}, and the other terms measured directly from KiDS-1000.  We note that the band encompasses the $2\sigma$ uncertainty from the measurement of $\alpha$.  All other error terms are sub-dominant.    

In contrast we find little agreement between the measured residual star-galaxy correlation function (shown pink), with the model prediction (grey dotted line, Eq.~\ref{eqn:gdemod}).    This suggests that our faint magnitude linear extrapolation of the residual PSF ellipticity, $\delta \epsilon^{\rm PSF}$, is not representative of the detector level bias that the galaxies have experienced.   We find that the average residual star-galaxy correlation is significantly lower than the expectation from the extrapolated chip-dependent residual PSF ellipticity model.  We therefore conclude that whilst we find significant flux dependence in the PSF residual ellipticity for 3/32 OmegaCAM chips (see Fig.~\ref{fig:chip_mag_residual}), there is no evidence in the shear catalogue that this leads to a significant bias in the cosmic shear measurements.  As such we conclude that there is no necessity for \citet{asgari/etal:2020b} to follow \citet{hildebrandt/etal:2020} in introducing a nuisance parameter in the fiducial KiDS-1000 cosmological parameter inference, to marginalise over this 2D residual PSF distortion.    For future surveys, with decreased statistical noise, this conclusion should, however, be reviewed.

Based on these results, we conclude that the linear systematics model provides a good representation of the systematics in our data, when $\delta \epsilon^{\rm PSF} = 0$.  As such we can use the \citet{bacon/etal:2003} systematics estimator $\xi_{\pm}^{\rm sys}(\theta)$ (Eq.~\ref{eqn:Baconsys}, analysing the corrected shear estimator $\epsilon^{\rm obs}_{\rm corr}$) to empirically estimate the systematic contribution to the measured cosmic shear signal.

\begin{figure}
\includegraphics[width=0.5\textwidth]{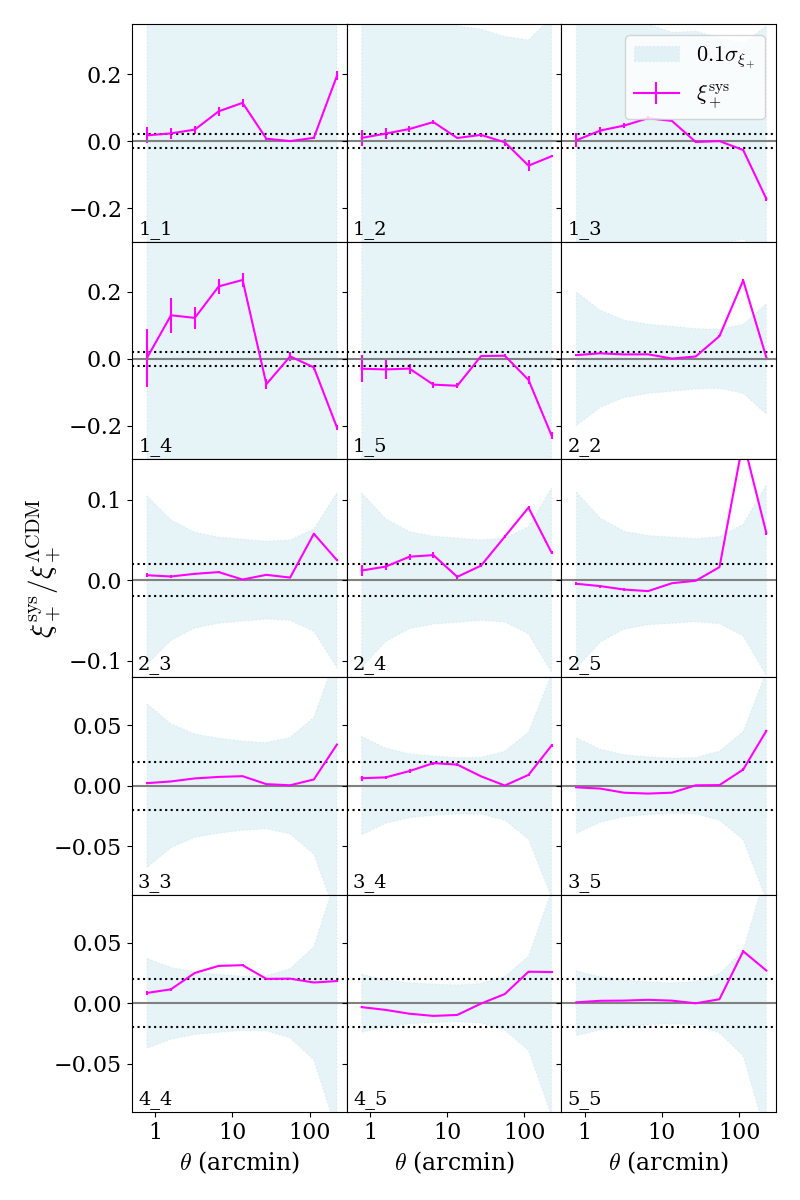} 
\caption{Ratio of the predicted systematic contribution to the expected amplitude of the cosmic shear signal, $\xi_+^{\rm sys}(\theta)/\xi_+^{\Lambda{\rm CDM}}(\theta)$, for 15 different tomographic bin combinations (as denoted in the lower left corner of each panel). The systematic contribution is typically either less than $\sim 2\%$ of the cosmic shear signal (shown as dashed lines), and/or within 10\% of the expected noise on the measurement (blue shaded region).}
\label{fig:Baconsys}
\end{figure}

In Fig.~\ref{fig:Baconsys} we compare the measured $\xi_+^{\rm sys}(\theta)$ for each tomographic bin combination with the amplitude of the expected fiducial cosmic shear signal, presenting the ratio of the two quantities\footnote{We note that 
we do not also show $\xi_-^{\rm sys}(\theta)$, as this is a very noisy quantity.  Both the numerator and denominator in Eq.~(\ref{eqn:Baconsys}) are essentially zero for the $\xi_-(\theta)$ estimator.}.   
This can also be compared to the expected noise in KiDS-1000, where we show $\pm10\%$ of the standard deviation of the cosmic shear signal \citep{joachimi/etal:2020} as a blue shaded region. In the majority of cases we find that the systematic bias remains within $\sim 10$\% of the statistical noise.   For the highest redshift bins, which carry the main cosmological constraining power for the survey, we typically find the systematic contribution to be less than $\sim 2\%$ of the cosmic shear signal (shown as dashed lines).    The exceptions are the large-scale, $\theta > 60$ arcmin, signal in some bins where the expected fiducial cosmic shear signal is small and the statistical noise is high. 

Using our rapid $\chi^2$ analysis (Eq.~\ref{eqn:chisqtest}) we assess the impact of this systematic by inspecting the goodness-of-fit of the fiducial cosmological model, given a $\xi_+^{\rm sys}(\theta)$ biased data vector.  We find the change in the goodness-of-fit to be consistent with the change expected when the data vector is instead drawn from a cosmological model where $S_8$ changes by $0.4\sigma_{S_8}$ compared to the fiducial case.   This systematic appears therefore to exceed our tolerance requirement of PSF modelling errors inducing less than a $0.1\sigma_{S_8}$ change in $S_8$. As such this systematic is flagged by our rapid $\chi^2$-test and referred to a complete bias impact assessment.  We follow \citet{troxel/etal:2018} by conducting a full MCMC cosmological inference analysis of a $\xi_{\pm}^{\rm sys}$ corrected KiDS-1000 cosmic shear data vector \citep[see][for details of the inference pipeline]{asgari/etal:2020b, joachimi/etal:2020}.   Comparing the resulting constraints on $S_8$ to the fiducial cosmic shear constraints we recover a bias of $0.06\sigma_{S_8}$ in the value of $S_8$.  This is well within our requirements and consistent with noise in the final converged MCMC chain \citep{joachimi/etal:2020}.   

To reconcile the two different conclusions from our rapid and full impact analysis, we remind the reader that they are testing different aspects of the analysis.  The $\chi^2$ test questions the goodness-of-fit of the model.  The MCMC analysis quantifies how the offsets found between the model and data transpire to bias the inferred cosmological parameter constraints.    For systematics that do not have the same angular or redshift scaling behaviour as the cosmological signal, the impact in terms of parameter bias is expected to be weak \citep[see for example][]{amara/etal:2008}.  For KiDS-1000, the changing sign in $\alpha$ from the fourth to the fifth bin results in a signal that adds to the auto-bins, and subtracts from the cross-correlation.   This type of behaviour predominantly impacts the intrinsic alignment modelling rather than the $\Lambda$CDM parameters.   Marginalising over the many different nuisance parameters in the MCMC analysis therefore allows for some degree of marginalisation over this systematic effect, albeit reducing the goodness-of-fit of the model which our $\chi^2$ test is based upon.     It is also worth noting that the MCMC inference analyses the full $\xi_{\pm}(\theta)$ data vector, where $\xi_{-}(\theta)$ is unaffected by this systematic, in contrast to our $\chi^2$ analysis from Eq.~(\ref{eqn:chisqtest}), which focuses on the impact from $\xi_{+}$ alone.    As the MCMC analysis provides a direct estimate of the $S_8$-bias introduced by the 
significant
 but low-level KiDS-1000 first-order systematics model, we conclude that the KiDS-1000 shear catalogue meets our current requirements.   
Future improvements to minimise $\alpha$ are however likely to be required as the statistical power of the survey increases.

\subsection{Comparison of the Paulin-Henriksson et al. and first-order systematic models}
 
Before concluding this section it is worth pausing to review the different systematic contributions to the two-point shear correlation function estimator as predicted by the PH08 model and the first-order systematics model.   Given that these two models have the same format (compare Eqs.~\ref{eqn:pherror} and \ref{eqn:H06error}), it may be tempting to link the different parameters where $m \equiv \overline{\delta T_{\rm PSF}/T_{\rm gal}}$, $\alpha \equiv \overline{\delta T_{\rm PSF}/T_{\rm gal}}$ and $\beta \equiv \overline{T_{\rm PSF}/T_{\rm gal}}$.   We find, however, that these quantities have very different amplitudes.   The empirical measurement of $\alpha$, shown in Fig.~\ref{fig:alphac}, and the shear calibration bias $m$, calibrated through image simulations, are both two orders of magnitude larger than the equivalent PH08 model terms.   

It is therefore important not to forget that the  PH08 model, also referred to in other studies as the `$\rho$-statistics'  \citep{rowe:2010}, was only ever intended to capture the contributions to the cosmic shear signal that arise from errors in the PSF modelling.   In the case of KiDS-1000, we find that the low-level PSF modelling errors are harmless in terms of the accuracy of the observed cosmic shear signal.   In contrast, however, there are other factors in the shear measurement that imprint PSF residual distortions and calibration biases in the shear estimator, 
such as object selection and noise bias 
\citep[see for example the discussion in][]{kannawadi/etal:2019}, in addition to weight-bias (discussed in Sect.~\ref{sec:lensfit}).   These factors are not captured by the PH08 model, and by using empirical estimates and image simulations we find that these factors are significant, adding to the cosmic shear signal at the level of a few percent.

We therefore recommend that the PH08 model is only used for its original intention, which is to optimise the functional form of the PSF model (as in Sect.~\ref{sec:psfncoeff}), and to validate the final PSF model.   For the validation of the shear catalogues,
 extra null-tests need to be undertaken, and our preferred approach is to adopt the linear systematics model with the parameters empirically determined from the catalogues.  Once the model is validated with the data 
 (see for example Fig.~\ref{fig:stargal}),
  the \citet{bacon/etal:2003} systematics estimator can then be used to determine the level of systematics that contribute to the observed cosmic shear signal.

\section{Two-point null-tests}
\label{sec:Null_tests}
In this section we extend our validation of the KiDS-1000 shear catalogue by presenting three additional two-point null-tests: analysis of B-modes, galaxy-galaxy lensing in the camera reference frame, and a shear-ratio test.

\subsection{COSEBIs B-modes}
\label{sec:bmodes}

\begin{figure}
  \begin{center}
    \includegraphics[width=0.5\textwidth]{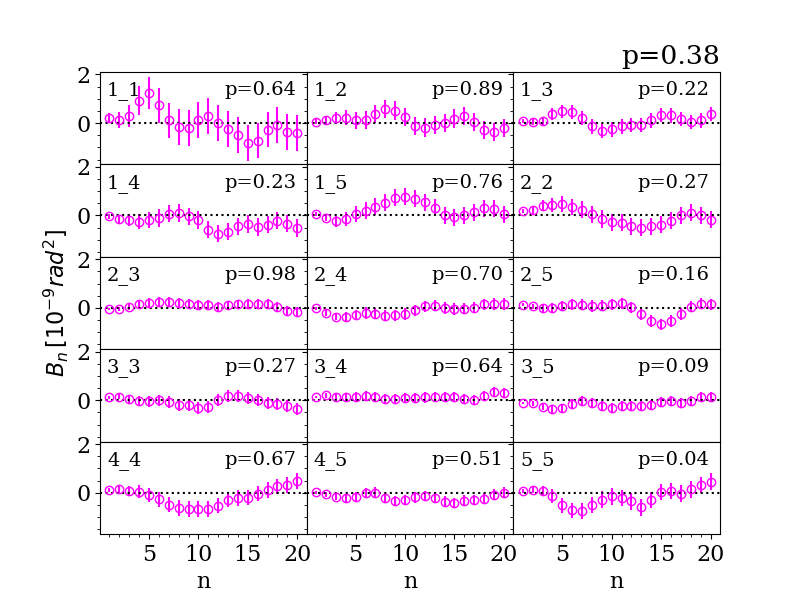}
    \end{center}
\caption{KiDS-1000 B-mode null-test:   the COSEBIs $B_n$ modes are
  shown for each tomographic bin combination (denoted in the upper
  left corner of each panel).   The measured B-modes are consistent with
  random noise, as determined through the $p$-values shown in the
  upper right corner of each panel.   Considering the full data vector of $n=20$ modes and 15
different tomographic bin combinations, we find $p=0.38$ corresponding to
an insignificant $0.3\sigma$ deviation from a null signal.   We
caution the reader against `chi-by-eye' as the $B_n$ modes are highly correlated. } 
\label{fig:Bmodes}
\end{figure}

Figure~\ref{fig:Bmodes} presents the B-mode signal, measured for each
tomographic bin combination, using Complete Orthogonal Sets of E/B
Integrals \citep[COSEBIs,][]{schneider/etal:2010}.    The COSEBIs
formalism allows for the clean and complete separation of the KiDS-1000 lensing
E-modes \citep[presented in][]{asgari/etal:2020b} from any
non-lensing B-modes.  It is therefore our preferred B-mode null-test
statistic \citep[see the discussion and comparison of B-mode statistics in][which concludes that the COSEBIs methodology provides the most
sensitive and stringent method to detect B-mode distortions]{asgari/etal:2019}.

The COSEBIs B-mode estimator is given by an integral over the two-point shear
correlation function, $\xi_\pm(\theta)$ from
Eq.~(\ref{eqn:twoptest}), as
\be
B_n = \frac{1}{2} \int_{\theta_{\rm min}}^{\theta_{\rm max}} {\rm
  d}\vartheta \, \vartheta \left[ T_{+n}(\vartheta)\xi_+(\vartheta) -
  T_{-n}(\vartheta)\xi_-(\vartheta) \right] \, ,
\label{eqn:Bn}
\ee
where the logarithmic COSEBI mode filter functions that we use, $T_{\pm n}$, are given in
equations 28 to 37 of \citet{schneider/etal:2010}.    We set
$\theta_{\rm min} = 0.5$ arcmin, and $\theta_{\rm max} = 300$ arcmin,
spanning the full angular range used in the cosmic shear analysis of \citet{asgari/etal:2020b}.
We measure $\xi_\pm(\theta)$ from Eq.~(\ref{eqn:twoptest}), using
4000 bins equally spaced in $\log \theta$, and calculate $B_n$ from
Eq.~(\ref{eqn:Bn}) by approximating the integral as a discrete sum
over the 4000 angular bins. Although it is tempting to analogise the COSEBI mode $n$ with the Fourier mode $\ell$, COSEBIs exist neither in Fourier nor real space, but in their own `COSEBI space' with each mode having contributions from a range of scales. $n$ should therefore be thought of as the index of the (generally oscillatory) filter functions used in converting the real-space $\xi_+(\vartheta), \xi_+(\vartheta)$ statistics to the COSEBI-space $B_n$ \citep[as well as its E-mode counterpart $E_n$; see equation 1 of][]{schneider/etal:2010}.

We assess the significance of the measured B-modes using an analytical
covariance matrix from \citet{joachimi/etal:2020}, that has been
verified through a series of mock simulation analyses, in addition to
an empirical `spin-test' whereby the shape-noise component of the
covariance is validated through numerous re-analyses of the survey
with randomised ellipticities \citep{troxel/etal:2018}.    We note
that `chi-by-eye' is dangerous with the COSEBIs statistic as
the modes are highly correlated.

We find that the measured B-modes, $B_n$, are consistent with zero with a
$p$-value\footnote{The $p$-value is equal to the probability of
  randomly producing a B-mode that is more significant than
  the measured B-mode signal, for the model that $B_n$ is drawn at random from a zero-mean Gaussian distribution.} of $p=0.38$ considering the full data vector of $n=20$ modes and 15
different tomographic bin combinations.   This corresponds to
an insignificant $0.3\sigma$ deviation from a null signal.   Analysing each
tomographic bin combination separately 
we find that all the B-modes are
consistent with zero, (see the $p$-values reported in the upper right
corner of each sub-panel in Fig.~\ref{fig:Bmodes}).   The lowest
$p$-value of $p=0.04$ is found for the
auto-correlation of the fifth tomographic bin, corresponding to a $1.8\sigma$ deviation from a null signal.    With 15 different
bin combinations, however, we do expect to find approximately one bin combination
with a $ \sim 2\sigma$ deviation from the null case.   We therefore
conclude that the measured KiDS-1000 B-modes are consistent with statistical noise.

\citet{asgari/etal:2012} show that the first $n=5$ E-modes
contain almost all the cosmological information.  It is therefore relevant to also limit the B-mode null-test to the first $n=5$ B-modes.   In this case 
our conclusions remain unchanged, finding $p=0.02$, which is consistent 
with zero B-modes at the $\sim 2\sigma$ level.   
Inspection of the 15 different bin combinations for the first $n=5$ B-modes, yields two $\sim 2\sigma$ deviations from the null case in the 2{\_}2 and 3{\_}5 combinations with $p = 0.02, 0.01$, respectively. For the 5{\_}5 bin combination highlighted 
as a potential outlier in the $n=20$ B-mode test, we note that $p = 0.23$, in the $n=5$ B-mode test.

\citet{asgari/etal:2019} demonstrate that some systematics influence the
E-modes and B-modes differently, and, as such, it is necessary to pass both these tests, which we do.   
Interestingly, they note that PSF residual distortions typically impact the low-$n$ modes.  The decrease 
in the $p$-values seen between the $n=20$ and the $n=5$ null-test, may therefore be a reflection of the
significant, but low-level, PSF residual distortions detected in Sect.~\ref{sec:phmodelinkids}.

\subsection{Galaxy-galaxy lensing in the OmegaCAM pixel reference frame}
Galaxy-galaxy lensing measures the azimuthally averaged tangential
shear of background galaxies relative to foreground galaxies
\citep{brainerd/etal:1996}.  In contrast to cosmic shear measurements,
where systematics increase the amplitude of the observed signal
(Eq.~\ref{eqn:H06error}), this azimuthal averaging typically
results in a cancellation of additive systematics.   As such, galaxy-galaxy
lensing is often regarded as a truly robust weak lensing probe \citep{mandelbaum/etal:2013}.

\begin{figure}
\begin{center}
\includegraphics[width=0.5\textwidth]{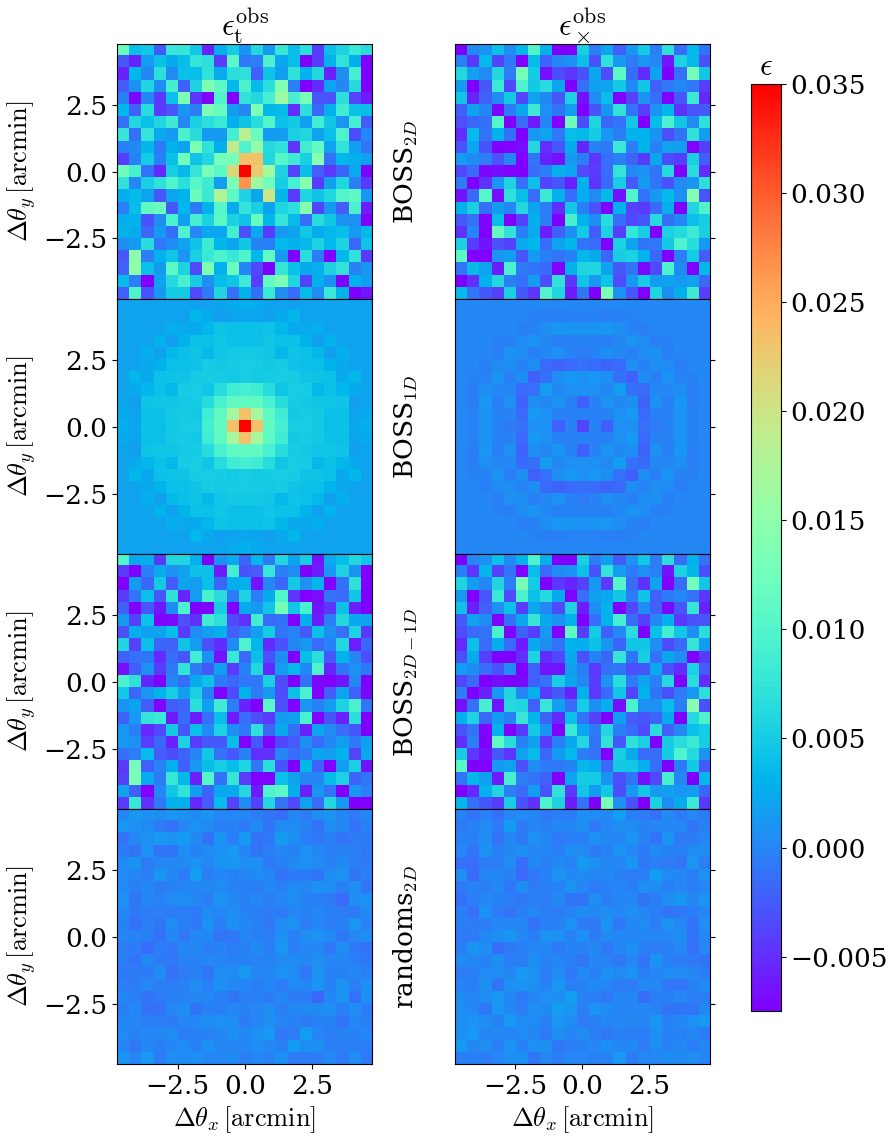} 
\caption{KiDS-BOSS galaxy-galaxy lensing in the OmegaCAM pixel
  reference frame for the tangential,   $\epsilon_{\rm t}^{\rm obs}$
  (left), and cross,  $\epsilon_\times^{\rm  obs}$ (right -
  null-test), components of the observed ellipticities in the fifth tomographic
  bin with $0.9 < z_{\rm B} < 1.2$. \textit{Upper:} 2D correlation functions
  centred on the BOSS lenses with $0.2 < z < 0.7$. \textit{Upper-middle:} azimuthally averaged 1D correlation
  functions around BOSS lenses, projected onto the 2D
  grid. \textit{Lower-middle:} the featureless residual 2D signal. \textit{Lower:} 2D correlation functions
  centred on random positions.} \label{fig:GGL_2D}
\end{center}
\end{figure}

In Fig.~\ref{fig:GGL_2D} we present the galaxy-galaxy lensing of
KiDS sources around foreground luminous red galaxies from BOSS
\citep{alam/etal:2015}, using the 409 deg$^2$
  of overlapping survey area between the equatorial KiDS-1000 region
  and BOSS.  We compare the standard one-dimensional (1D) azimuthally averaged
measurement with the signal measured on a 2D grid, within the
reference frame of OmegaCAM\footnote{All measurements are made using
  {\sc TreeCorr} \citep{jarvis/etal:2004,jarvis:2015}, with the 2D
  measurement facilitated by the \Verb:bin_type=TwoD: mode.}.
As
expected, the 2D measurement (upper panels) is noisier than the 1D measurement
(second panels).  We can use the residual between these
two measurements, however, to search for new systematic errors (third panel), 
finding no significant features.

This result is consistent with expectations from
the low-level systematics 
detected in Sect.~\ref{sec:PSF_Estimation}.  As a new null-test, it
does however allow us to explore
alternative systematics that are specific to galaxy-galaxy lensing
studies.   The featureless 2D residuals allow us to rule out any significant
differences in the behaviour of systematic errors in regions near bright objects, \citep[see for
example the discussions in][]{sheldon/huff:2017,sifon/etal:2018}.   We can also rule
out any significant impact from detector level defects, such as charge transfer
inefficiencies, in the region of the bright BOSS galaxies.   This
analysis therefore
provides additional confidence in the standard `1D' azimuthally averaged galaxy-galaxy lensing
measurements that are part of our joint KiDS-BOSS multi-probe
lensing and clustering cosmological constraints presented in \citet{heymans/etal:2020}.

Fig.~\ref{fig:GGL_2D}
presents the result for the
fifth tomographic bin, as this has the strongest PSF contamination
fraction $\alpha$ (see Fig.~\ref{fig:alphac}), out to a maximum
radius of 5 arcmin from the central BOSS galaxy.   Our conclusions are
unchanged when analysing each of the other tomographic bins or the
full source sample, and when extending the analysis to 30 arcmin.
We remind the reader that we have empirically corrected the source catalogue
ellipticities
(see Sect.~\ref{sec:cterm}).
Given the featureless residuals (third panels), and the
featureless 2D signal measured around random points (lower panels), we
conclude that our approximation that the additive $c$-term is constant across
the survey, is also appropriate in the regions around bright galaxies.
    
 \subsection{Galaxy-galaxy shear-redshift scaling}
 \label{sec:Shear_ratio}   

For a fixed lens population, the ratios of the azimuthally averaged tangential shear, $\gamma_{\rm t}$, from different source populations is sensitive only to the ratios of the angular diameter distances between the source and lens planes. This `shear-ratio test' was originally conceived as a cosmological probe through the distance-redshift relation \citep{jain/taylor:2003}, although it was subsequently found that the cosmological dependence is rather weak \citep{taylor/etal:2007}.   This approach was therefore proposed as a unique joint null-test of the accuracy of the estimated source galaxy redshift distributions and the redshift-dependent shear calibration correction \citep{hoekstra/etal:2005,heymans/etal:2012}.

\subsubsection{The shear-ratio test: Measurement and modelling} \label{subsubsec:SRT}

Adopting an isolated singular isothermal sphere (SIS) as a model for the lens density profile, the mean tangential shear within an annulus of angular size $\theta$, $\gamma_{\rm t, SIS}(\theta)$, is given by
\begin{equation}
\gamma_{\rm t, SIS}^{ij}(\theta) = \frac{2\pi}{\theta} \left( \frac{\sigma^i_{\rm v}}{c} \right)^2 \beta^{ij} \,,
\label{eqn:sigv}
\end{equation}
where $\sigma^i_{\rm v}$ is the velocity dispersion of the lens galaxies in bin $i$, and $j$ denotes the source redshift bin \citep{bartelmann/schneider:2001}. The lens-averaged lensing efficiency, $\beta^{ij}$ is given by
\begin{equation}
\beta^{ij} = \int_0^\infty {\rm d}z_{\rm l} \, n^i(z_{\rm l}) \, \int_{z_{\rm l}}^\infty {\rm d}z_{\rm s} \, n^j(z_{\rm s}) \frac{D(z_{\rm l},z_{\rm s})}{D(0,z_{\rm s})} \,,
\end{equation}
where $D(z_{\rm a}, z_{\rm b})$ is the angular diameter distance between redshifts $z_{\rm a}$ and $z_{\rm b}$, and $n^i(z)$ and $n^j(z)$ are the redshift distributions of the lenses and sources, respectively.  The recovery of a consistent set of constraints on the velocity dispersion $\sigma^i_{\rm v}$ using a range of different tomographic source bins, $j$, serves as a validation of the shear and redshift estimates \citep{heymans/etal:2012}.    

There are four key assumptions made when using Eq.~(\ref{eqn:sigv}) to model measurements of tangential shear around lens galaxies. 
In this equation the lens is assumed to have a perfect SIS density profile.   It is considered isolated, an assumption which is only reasonable to make on small scales \citep[see for example][]{velander/etal:2014}.  Any intrinsic alignment (IA) terms between the source and lens galaxies are neglected \citep[see for example][]{joachimi/etal:2015}.  The weak lensing magnification of background sources by the matter associated with the foreground lenses is also neglected \citep{unruh/etal:2019}. 

To address the issue of the choice of lens model, we follow \citet{hildebrandt/etal:2017} who 
developed the shear-ratio test such that it was agnostic to the galaxy halo density profile by modelling the tangential shear $\gamma_{\rm t}^{ij}(\theta) =A^i(\theta)\, \beta^{ij}$.   Here $A^i(\theta)$ is a $\theta$-dependent set of free parameters for each lens bin $i$ \citep[see also][who choose to model $A^i(\theta)$ as a power law]{prat/etal:2018}.     

\begin{figure*}
\includegraphics[width=\textwidth]{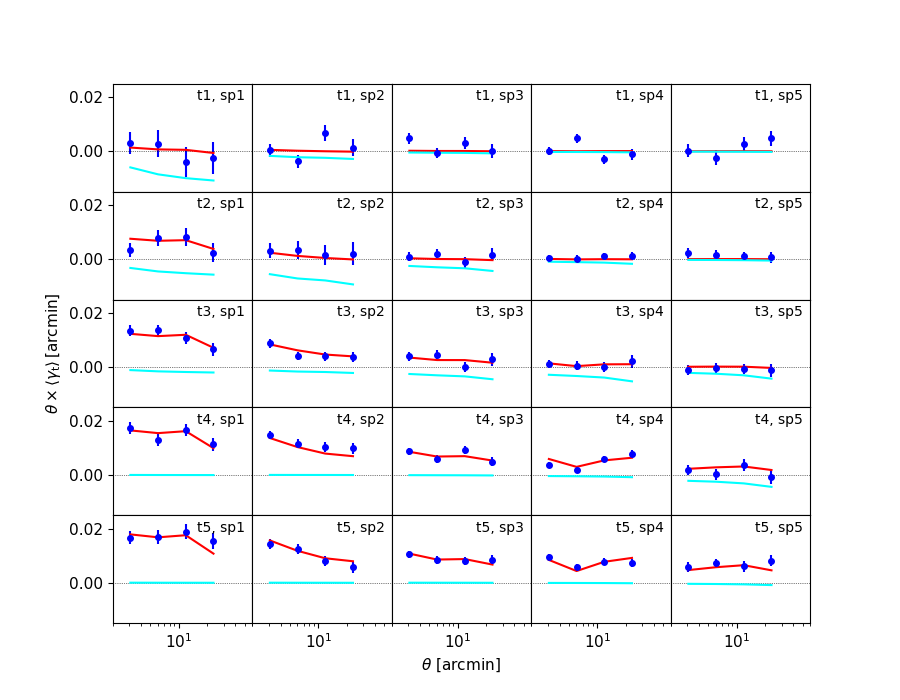} \\
\caption{Azimuthally averaged tangential shear around BOSS galaxies (blue data points) for each tomographic (labelled `t 1-5') and spectroscopic (`sp 1-5') redshift bin combination. These can be compared with the best-fit $A^i(\theta)$ with IA model (red lines) which is consistent with the data at the $1.6\sigma$ level.  The cyan lines display the predicted contribution to the tangential shear model from intrinsic alignments, scaled by a factor of five to aid visualisation.}
\label{fig:SRT}
\end{figure*}

To address the IA terms in the observed signal, our fiducial analysis includes the `NLA' intrinsic alignment model from \citet{bridle/king:2007} with a fixed IA amplitude $A_{\rm IA}=1.0$, and a linear galaxy bias of $b=2.0$, which are reasonable amplitudes for the KiDS source and BOSS lens populations that we study \citep{joachimi/etal:2020}.    For this model we find that the IA contribution is non-negligible for the source-lens combinations where there is significant overlap between the source and lens samples (see the cyan lines in Fig.~\ref{fig:SRT}).  As such it is important to include this additional IA signal in our null-test.  Our adopted model for the shear-ratio test is therefore given by
\be
\gamma_{\rm t}^{ij}(\theta) =A^i(\theta)\, \beta^{ij} - A_{\rm IA} \int_0^{\infty} \frac{\ell {\rm d}\ell}{2\pi} \,{\rm J}_2(\ell\theta) \, C_{\rm gI}^{ij}(\ell) \, ,
\label{eqn:SRTmod}
\ee
where ${\rm J}_2$ is the second order Bessel function of the first kind, and $C_{\rm gI}(\ell)$ denotes the angular power spectrum of the intrinsic ellipticity alignment of source galaxies, that are physically close to the lenses\footnote{This term is defined in, for example, equation 24 of \citet{joachimi/etal:2020}.  We note, however, that we have taken the scaling factor of $-A_{\rm IA}$ out to the front of the integral in order to clarify that $C_{\rm gI}(\ell)$ scales linearly with this free parameter, and that it serves to reduce the amplitude of the observed signal.}.

To explore our sensitivity to weak lensing magnification bias, and in order to verify our methodology, we analyse mock KiDS and BOSS galaxy catalogues constructed from the MICE2 simulation 
\citep{fosalba/etal:2015,hoffman/etal:2015,carretero/etal:2015,crocce/etal:2015,fosalba/etal:2015b}
using the pipeline\footnote{The \citet{vandenbusch/etal:2020} KiDS-MICE mock catalogue pipeline can be downloaded from \url{https://www.github.com/KiDS-WL/MICE2_mocks.git}} from \citet{vandenbusch/etal:2020}.  MICE2 is based on an N-body dark
matter simulation, which is used to derive an all-sky lensing mock catalogue between $0.1\leq z \leq 1.4$, along with mock galaxy catalogues.   These catalogues are sampled to 
carefully match the properties of KiDS and BOSS, including their redshift distributions, overlap and sample selection \citep[see][for details]{wright/etal:2019,vandenbusch/etal:2020}.    As MICE allows for the inclusion or exclusion of weak lensing magnification, this mock also allows us to quantify the impact of neglecting magnification in our shear-ratio model.  For the source and lens samples used in this analysis, we find that the magnification bias is sufficiently small to be considered negligible given the signal-to-noise of our analysis, changing the amplitude of the galaxy-galaxy lensing signal by $(0.04 \pm 0.03)\, \sigma_{\gamma_{\rm t}}$, per bin, where $\sigma_{\gamma_{\rm t}}$ is the measured error for a $\theta$ and source-lens bin.   Whilst the effect of IAs and magnification on the tangential shear are comparable in size, the latter contributes most strongly at scales where the galaxy-galaxy lensing signal itself is large.  Magnification therefore has a significantly smaller relative impact on our shear-ratio null-test, compared to IAs.   
  
With our model in place, we analyse 
the galaxy-galaxy lensing signal around luminous red galaxies from the BOSS spectroscopic survey \citep{alam/etal:2015}, divided into five narrow redshift bins of width $\Delta z=0.1$ between $0.2 \leq z \leq 0.7$, for the five KiDS-1000 tomographic bins in Table~\ref{tab:catstats}.   The lens bins were chosen to be sufficiently narrow to minimise galaxy bias evolution across the bin, but also sufficiently broad to produce a reasonable signal-to-noise null-test.  We adopt the \citet{mandelbaum/etal:2005} galaxy-galaxy lensing estimator,
\eqa{
\label{eqn:ggl_estimator}
\widehat{\gamma_{\rm t}}(\theta) = \frac{1}{1+m_{\rm s}}  & \left( \, \frac{\sum_{\rm ls} w_{\rm l}\, w_{\rm s}\, \epsilon_{{\rm t,l} \rightarrow {\rm s}}\, \Delta_{\rm ls}(\theta)}{\sum_{\rm rs} w_{\rm r}\, w_{\rm s}\, \Delta_{\rm rs}(\theta)}  {\cal N}_{\rm rnd}   \right. \\ \nn
& \left.  - \, \frac{\sum_{\rm rs} w_{\rm r}\, w_{\rm s}\, \epsilon_{{\rm t,r} \rightarrow {\rm s}}\, \Delta_{\rm rs}(\theta)}{\sum_{\rm rs} w_{\rm r}\, w_{\rm s}\, \Delta_{\rm rs}(\theta)} \, \right) \, ,
}
where $m_{\rm s}$ is the shear calibration correction for source bin $s$, $\epsilon_{\rm {t,l} \rightarrow {\rm s}}$ is the tangential shear measured around lenses, $\epsilon_{\rm{t,r} \rightarrow {\rm s}}$ is the tangential shear measured around random points within the BOSS footprint, $\Delta_{}(\theta)$ is the angular binning function (see Eq.~\ref{eqn:twoptest}), and the sums with indices ${\rm l}$, ${\rm s}$, and ${\rm r}$ run over all objects in the lens, source, and random catalogues, respectively.   The normalisation term ${\cal N}_{\rm rnd} := \sum_{\rm r} w_{\rm r}/ \sum_{\rm l} w_{\rm l}$ reduces to the oversampling factor of the random catalogue with respect to the catalogue of lens galaxies, for unit weights.  In this analysis we use roughly 100 times as many random points as lenses.  Finally, the weights consist of the source {\it lens}fit weights $w_{\rm s}$, and the BOSS completeness weights for the galaxy sample, $w_{\rm l}$, and random catalogue, $w_{\rm r}$, which have unit value for all random points.      

The galaxy-galaxy lensing estimator in Eq.~(\ref{eqn:ggl_estimator}) automatically corrects for dilution-effects arising from the source galaxies that are clustered with the lens\footnote{This can be seen by recasting the first term in Eq.~(\ref{eqn:ggl_estimator}) as the simple $\gamma_{\rm t}$ estimator with $\gamma_{\rm t}(\theta)= (\sum_{\rm ls} w_{\rm l}\, w_{\rm s} \epsilon_{{\rm t}})/{\sum_{\rm ls} w_{\rm l}\, w_{\rm s}}$ scaled by the ratio between the weighted number of galaxy pairs in the source-lens, ${\sum_{\rm ls} w_{\rm l}\, w_{\rm s}}$,  and source-random sample ${\sum_{\rm rs} w_{\rm r}\, w_{\rm s}}$, modulo the normalisation term ${\cal N}_{\rm rnd}$.   In the case where the source and lens samples are unclustered, for example at large angular separations, this normalised ratio is unity, and the estimator in Eq.~(\ref{eqn:ggl_estimator}) returns to the simple format.   In the case where the sources and lenses physically cluster, the effective pair count is higher in the source-lens sample than in the source-random sample.   The inclusion of this term therefore boosts the simple $\gamma_{\rm t}$ signal by a factor that accounts for the small fraction of sources that are physically connected to the lens and are thus diluting the overall signal.    With this correction, the effective redshift distribution of the sources is given by the average source redshift distribution at each angular scale.}.  Here the angular dependence of the clustering of galaxies modifies the average redshift distribution of source galaxies as a function of their angular separation from the lens, with close-separation source-lens pairs more likely to be sampled from the source $n(z)$ at the location of the lens \citep[see for example][]{hoekstra/etal:2015}.     By also including a `random correction', the second term in Eq.~(\ref{eqn:ggl_estimator}), we reduce the sampling noise terms that arise from the large-scale structure \citep{singh/etal:2017}.  We also reduce the shape noise terms that arise from the different fraction of unique sources used in each $\theta$-bin \citep[see figure E.1 in][]{joachimi/etal:2020}.   We follow \citet{hildebrandt/etal:2017} by using four angular scales logarithmically spaced between 2 and 30 arcminutes, where the scales were chosen to minimise the amplitude of these boost and random correction terms \citep[see][for further discussion on these points]{blake/etal:2020}.  This results in 100 data points (four per lens-source bin, with 25 lens-source bin combinations), to which we simultaneously fit a 20-parameter $A^i(\theta)$ model (one per lens bin $i$, with four $\theta$ scales).

One of the most challenging aspects for this null-test is the determination of an accurate covariance matrix, as the Limber equation is an approximation that becomes less accurate as the width of our lens bins narrow \citep{giannantonio/etal:2012, kilbinger/etal:2017}.  This precludes the use of fast Limber approximated analytical covariances \citep[although see][for new efficient beyond-Limber calculations to mitigate this issue in the future]{fang/etal:2020}.   
Even with the MICE2 simulation's angular size spanning an octant of the sky $(\sim 5000 \, \rm{deg}^2)$, a single realisation provides insufficient area to construct a covariance matrix from these mocks that can be accurately inverted. We therefore follow \citet{troxel/etal:2018} in determining a covariance from 500 `spin' realisations of the shear field, whereby each source galaxy is randomly rotated, resulting in a nulled, noisy signal.   This approach results in a conservative covariance estimator that does not include sampling variance, rendering the shear-ratio test a more challenging test to pass. We note however that for the $\theta$-scales used in our analysis, the sampling variance terms are expected to be sub-dominant \citep[see][for details]{joachimi/etal:2020}.

Figure~\ref{fig:SRT} presents the KiDS-BOSS galaxy-galaxy lensing measurements between five spectroscopic lens bins (left to right), and five tomographic source bins (upper to lower).  
The measurements can be compared to the best-fit model from Eq.~(\ref{eqn:SRTmod}) 
(shown red), calculated using the redshift distributions of the SOM-gold sample described in Sect.~\ref{subsec:photoz}. 
We calculate the $\chi^2$ goodness of fit, and recast this as a $p$-value which describes the probability of the data being drawn from the model. We find that the 5-bin model provides a reasonable fit to the data, with $p=0.050$, such that the 5-bin null-test is consistent with the model expectation at the $1.6\sigma$ level\footnote{We note that when conducting a similar analysis using spectroscopically confirmed lens galaxies from GAMA, we find a good fit of the 5-bin model to the data, with $p=0.185$,
 confirming the 
 KV450 
 analysis of \citet{hildebrandt/etal:2020}.   This is expected as the overlap between KiDS-1000 and GAMA is unchanged from the previous KiDS data release.   \citet{hildebrandt/etal:2020} also find a reasonable model fit for a shear-ratio test conducted with KV450 and BOSS.   With the overlapping KiDS-BOSS area doubling in KiDS-1000, our KiDS-BOSS shear-ratio null-test is now more constraining, improving our ability to detect any inconsistencies in our data set.}.   We recognise that the three highest redshift bins contribute the majority of the cosmological constraining power for KiDS-1000.   It is therefore also prudent to conduct a shear-ratio test for these three bins alone.   In this 3-bin null-test we find a good fit to the data with $p=0.299$, such that the 3-bin null-test is consistent with the model expectation at the $0.5\sigma$ level. With this degree of consistency, for both the full 5-bin null-test and the restricted 3-bin null-test, we conclude that 
the
KiDS-1000 joint shear and redshift calibration has passed this final null-test.  
We refer the reader to \citet{asgari/etal:2020b}, who carry out a complementary cosmic shear consistency test for the different tomographic bins following the methodology of \citet{kohlinger/etal:2019}.

\subsubsection{The shear-ratio test: Sensitivity to shear-redshift calibration errors and the intrinsic alignment model} \label{subsubsec:SRT_sensitivity}
In order to establish the sensitivity of the KiDS-1000 shear-ratio test to shear and redshift calibration errors we determine $p$-values for a series of test cases.   We start with our fixed fiducial intrinsic alignment model, and coherently bias the mean estimated redshift of each tomographic bin with an increase of $5\sigma_z$, where $\sigma_z$ is given in Table~\ref{tab:catstats}.  In this case we find $p=0.024$.  Coherently decreasing the mean estimated redshift of each tomographic bin by $5\sigma_z$, 
we find $p=0.047$.  Constructing an incoherent model, where the estimated $n(z)$ are biased alternately by $\pm 5\sigma_z$ we find $p=0.016$.    Turning to the shear calibration correction, $m$, we increase the calibration correction by $\pm 5\sigma_m$ in alternating bins, where $\sigma_m$ is given in Table~\ref{tab:catstats}.  In this case we find $p=0.012$.      

From these tests one can conclude that the shear-ratio test is fairly insensitive to calibration errors in the shear and mean redshift at less than the $5\sigma$ level in the case of a known intrinsic alignment model.
The inability of the shear-ratio test to discriminate between these systematics, given the level of statistical uncertainty in KiDS-1000, means that although our data is shown to meet the null-test requirements (see Sect.~\ref{subsubsec:SRT}), this fact does not necessarily eliminate the prospect of such biases residing in our data.

In \citet{asgari/etal:2020b} the uncertainty on the amplitude of the intrinsic alignment model $A_{\rm IA}$ is accounted for by marginalising over $A_{\rm IA}$ with an uninformative top-hat prior ranging from $-6 < A_{\rm IA} < 6$.   We estimate the impact of marginalising over this level of uncertainty in our shear-ratio test by increasing the errors on $\gamma_{\rm t}(\theta)$ by a factor given by the expected IA signal with $A_{\rm IA}=6$.   In this case our fiducial analysis passes with $p=0.913$.   We can also adopt a more realistic, but informative prior with $-2 < A_{\rm IA} < 2$, motivated by the $\pm 3\sigma$ constraints on $A_{\rm IA}$ in \citet{wright/etal:2020b}.  In this case our fiducial analysis passes with $p=0.179$.   When including an uncertainty in the intrinsic alignment model, we find all our $5\sigma$ test sensitivity cases, as described above, pass with $p>0.033$. 

We pause to note that marginalising over the uncertainty in the IA model is particularly relevant for the first two tomographic bins, which, when analysed alone present a moderate tension between the data and model expectation with $p=0.009$. With the inclusion of the IA model uncertainty, the shear-ratio test for the first two tomographic bins alone pass with $p=0.024$. 

This study demonstrates that in order to exploit shear-ratio observations to provide a precise validation of the calibration of shear and redshift estimates, 
these small $\theta$-scale galaxy-galaxy lensing observations must be used in 
conjunction with other cosmological probes in order to constrain the intrinsic alignment terms in the model \citep{maccrann/etal:2020}.  A joint simultaneous analysis also removes the necessity to assume that the shear-ratio test is insensitive to cosmology, as this assumption is further challenged by the introduction of intrinsic-alignment modelling. Furthermore, the minimal impact of shear and redshift estimation biases on the results of the shear-ratio test demonstrated here, is evidence of the limitations of this test to validate weak lensing catalogues given the statistical power of the current stage-III surveys.

\section{Conclusions and summary}
\label{sec:Conc}
In this analysis we have presented the shear catalogues for the fourth data release of the Kilo-Degree Survey\footnote{The KiDS-1000 shear catalogue is publicly available at \href{http://kids.strw.leidenuniv.nl/DR4/lensing.php}{kids.strw.leidenuniv.nl/DR4/lensing.php}}, KiDS-1000.   This survey spans 1006 square degrees with high-resolution deep imaging down to $r= 25.02 \pm 0.13$ (5$\sigma$ limiting magnitude in a 2 arcsec aperture with a mean seeing of 0.7 arcsec).  Over a total effective area of 777.4 square degrees, accounting for the area lost to multi-band masks, KiDS-1000 is fully imaged in nine bands with matched depths that span the optical to the NIR ($ugriZYJHK_{\rm s}$).   KiDS overlaps with a wide range of complementary spectroscopic surveys.  We additionally observe 4 square degrees of matched nine-band imaging targeting deep spectroscopic survey fields 
outside
the KiDS footprint.  KiDS-1000 therefore represents a unique survey of large-scale structure, owing to its design that mitigates two of the greatest challenges in weak lensing studies.   These are accurate shear measurements,
 facilitated by
 high signal-to-noise, high-resolution and stable imaging, 
 as well as 
 accurate photometric redshift estimation, aided by the extended wavelength coverage and extensive calibration fields \citep{wright/etal:2020}.     
The different trade-offs between area covered, depth attained, image quality delivered and wavelength range covered make KiDS-1000 nicely complementary to the other concurrent Stage-III weak lensing surveys, DES and HSC. With the development and application of a myriad of analysis tasks and tools, required to realise robust cosmological information from pixel-level imaging, the three-pronged approach of these independent Stage-III teams optimally serves the cosmological community in the final phases before the next generation of `full-sky' imaging surveys see first light over the coming few years.

This paper presents a series of null-tests to verify the robustness of the KiDS-1000 shear measurements, estimated using the model-fitting pipeline {\it lens}fit \citep{miller/etal:2013, fenechconti/etal:2017,kannawadi/etal:2019}.    The developments since the previous KiDS-450 release focus on upgrading the star selection for PSF modelling in Sect.~\ref{sec:stargalsep}, PSF model optimisation in Sect.~\ref{sec:psfncoeff}, detailed studies of detector level effects in Sect.~\ref{sec:detectorlevel}, and improvements in our weight bias correction in Sect.~\ref{sec:lensfit}.     

We review two approaches to set requirements on the accuracy of a shear catalogue.   
The \citet[PH08,][]{paulin-henriksson/etal:2008} systematics model, also referred to in other studies as the $\rho$-statistics \citep{rowe:2010},  
captures the contributions to the cosmic shear signal that arise from errors in the PSF modelling.   Using a simple $\chi^2$ test to quantify the impact of the inferred systematic contributions, in Sect.~\ref{sec:requirements}, we find that the accuracy of the KiDS-1000 PSF model is well within our requirements of not introducing more than a $0.1\sigma_{S_8}$ change in the recovered cosmological parameter $S_8 =\sigma_8\sqrt{\Omega_{\rm m}/0.3}$.  This $0.1\sigma$ limit is the typical variance between different MCMC parameter inference analyses using different random seeds \citep{joachimi/etal:2020}.   As the 
PH08 test does not capture other factors in the shear measurement that imprint PSF residual distortions and 
calibration biases in the shear estimator, however, we therefore also review a linear systematics model in Sect.~\ref{sec:H06sysmod} where the parameters are estimated empirically from the catalogues.   We verify that this model provides a suitable description of the systematics in the KiDS-1000 catalogues through a series of one-point and two-point consistency tests, enabling the use of the \citet{bacon/etal:2003} estimator to determine the resulting systematic contribution to the cosmic shear signal.  Our simple $\chi^2$ test, which quantifies the change in the goodness of fit of the cosmological model to the observations, raises a flag at the level of systematics identified in this analysis.  To quantify the bias that these systematics introduce in the inferred cosmological model, we conduct a full MCMC analysis of the KiDS-1000 cosmic shear data vector with the modelled systematic correction applied.  As the shape and amplitude of the detected PSF residual systematics across the tomographic bin combinations are sufficiently different from the behaviour of the cosmological parameters \citep[see for example the discussion in][]{amara/etal:2008}, we find that the systematic-corrected analysis differs by less than $0.06\sigma_{S_8}$ from the fiducial analysis.  From this we conclude that for the statistical noise levels in the KiDS-1000 data, the low-level PSF-residual systematics that we uncover in our analysis make a negligible impact on our cosmological parameter constraints.    This is supported by our B-mode analysis 
in Sect.~\ref{sec:bmodes}, where we decompose the signal into its cosmological E-mode and non-lensing B-modes, finding the B-modes to be consistent with pure statistical noise.

Our final null-test scrutinises both the shear and photometric redshift estimates, finding consistent constraints on the properties of BOSS luminous red galaxies from a series of different tomographic source samples.   This `shear-ratio' test, in Sect.~\ref{sec:Shear_ratio} simultaneously validates the redshift-dependent shear calibration correction from \citet{kannawadi/etal:2019} and the self-organising map photometric redshift calibration from \citet{wright/etal:2019}. We recognise that the shear-ratio null-test, where low photometric-redshift source galaxies are placed in front of high-redshift lenses, currently provides our primary test of the accuracy of the high-redshift $z>1.4$ tail of the photometric redshift distributions.   The $z>1.4$ galaxy sample is a redshift regime that we are currently unable to examine through other routes owing to a lack of mock galaxy catalogues that reliably extend galaxy colours beyond $z>1.4$ \citep{fosalba/etal:2015, derose/etal:2019}, and a lack of signal-to-noise in our cross-correlation analysis \citep{hildebrandt/etal:2020b}.  In this analysis, however, we find that when accounting for the contribution to the signal from the intrinsic alignment of galaxies, without a strong prior on the amplitude of the intrinsic alignment model,  the shear ratio test becomes significantly less sensitive to biases in the redshift or shear calibration. The \citet{maccrann/etal:2020} proposal to incorporate the small $\theta$-scale shear-ratio test into a multi-probe data vector for cosmological inference analyses therefore presents a promising route forward for 
upcoming studies.  
Future work to calibrate the $z_{\rm B} > 1.2$ galaxy sample in KiDS also necessitates the analysis of new high-redshift mock galaxy catalogues \citep[for example the Euclid Flagship simulations from][]{potter/etal:2017}.

KiDS completed survey observations in July 2019, spanning 1350 square degrees of imaging with a second pass in the $i$-band to facilitate a long-mode transient study.   Additional survey time was awarded to expand the overlap of nine-band imaging with deep spectroscopic surveys.   This includes $\sim 12$ square degrees targeting the VIPERS fields \citep{guzzo/etal:2014} and $\sim 4$ square degrees targeting additional VVDS fields \citep{lefevre/etal:2013}.  We therefore look forward to the fifth and final release and analysis of the ESO public KiDS, along with new results from 
DES and HSC, 
as well as 
the first-light imaging from the upcoming \textit{Euclid} survey and the Vera C. Rubin Observatory Legacy Survey of Space and Time.

\begin{acknowledgements}
We thank Mike Jarvis for his continuing enhancements, clear		
 documentation and maintenance of the excellent {\sc TreeCorr} software package, our external blinder Matthias Bartelmann 
held the key for which of the three catalogues analysed was the true unblinded catalogue, a secret revealed at the end of the KiDS-1000 study, and the anonymous referee for their useful comments and positive feedback on this paper.   We also wish to thank the Vera C. Rubin Observatory LSST-DESC Software Review Policy Committee (Camille Avestruz, Matt Becker, Celine Combet, Mike Jarvis, David Kirkby, Joe Zuntz with CH) for their draft Software Policy document which we followed, to the best of our abilities, during the KiDS-1000 project.   Following this policy, the software used to carry out the various different null-test analyses presented in this paper is open source at \href{https://github.com/KiDS-WL/Cat_to_Obs_K1000_P1}{github.com/KiDS-WL/Cat\textunderscore to\textunderscore Obs\textunderscore K1000\textunderscore P1}.    
 None of the KiDS-1000 papers would have been possible without the fantastic HPC support that we've gratefully received from Eric Tittley at the IfA. 
We are also very grateful to Elena Sellentin and Mohammadjavad Vakili for useful discussions and helpful comments whilst drafting the paper.\\
 
This project has received funding from the European Union's Horizon 2020 research and innovation programme: 
We acknowledge support from the European Research Council under grant agreement No.~647112 (BG, CH, MA, CL and TT),  and
No.~770935 (HHi, JLvdB, AHW and AD). 
BG also acknowledges the support of the Royal Society through an Enhancement Award (RGF/EA/181006). 
CH also acknowledges support from the Max Planck Society and the Alexander von Humboldt Foundation in the framework of the Max Planck-Humboldt Research Award endowed by the Federal Ministry of Education and Research. 
TT also acknowledges support under the Marie Sk\l{}odowska-Curie grant agreement No.~797794. 
HHi is also supported by a Heisenberg grant of the Deutsche Forschungsgemeinschaft (Hi 1495/5-1). 
HHo and AK acknowledge support from Vici grant 639.043.512, financed by the Netherlands Organisation for Scientific Research (NWO). 
KK acknowledges support by the Alexander von Humboldt Foundation. 
MB is supported by the Polish Ministry of Science and Higher Education through grant DIR/WK/2018/12, and by the Polish National Science Center through grants no. 2018/30/E/ST9/00698 and 2018/31/G/ST9/03388. 
JdJ is supported by the Netherlands Organisation for Scientific Research (NWO) through grant no. 621.016.402. 
NRN acknowledges financial support from the “One hundred top talent program of Sun Yat-sen University” grant no. 71000-18841229. 
HYS acknowledges support from NSFC of China under grant 11973070, the Shanghai Committee of Science and Technology Grant no.19ZR1466600 and Key Research Program of Frontier Sciences, CAS, grant no. ZDBS-LY-7013. \\

The results in this paper are based on observations made with ESO Telescopes at the La Silla Paranal Observatory under programme IDs 177.A-3016, 177.A-3017, 177.A-3018 and 179.A-2004, and on data products produced by the KiDS consortium. The KiDS production team acknowledges support from: Deutsche Forschungsgemeinschaft, ERC, NOVA and NWO-M grants; Target; the University of Padova, and the University Federico II (Naples). Contributions to the data processing for VIKING were made by the VISTA Data Flow System at CASU, Cambridge and WFAU, Edinburgh. \\

The BOSS-related results in this paper have been made possible thanks to SDSS-III. Funding for SDSS-III has been provided by the Alfred P. Sloan Foundation, the Participating Institutions, the National Science Foundation, and the U.S. Department of Energy Office of Science.   SDSS-III is managed by the Astrophysical Research Consortium for the Participating Institutions of the SDSS-III Collaboration including the University of Arizona, the Brazilian Participation Group, Brookhaven National Laboratory, Carnegie Mellon University, University of Florida, the French Participation Group, the German Participation Group, Harvard University, the Instituto de Astrofisica de Canarias, the Michigan State/Notre Dame/JINA Participation Group, Johns Hopkins University, Lawrence Berkeley National Laboratory, Max Planck Institute for Astrophysics, Max Planck Institute for Extraterrestrial Physics, New Mexico State University, New York University, Ohio State University, Pennsylvania State University, University of Portsmouth, Princeton University, the Spanish Participation Group, University of Tokyo, University of Utah, Vanderbilt University, University of Virginia, University of Washington, and Yale University.\\

The MICE simulations have been developed at the MareNostrum supercomputer (BSC-CNS) thanks to grants AECT-2006-2-0011 through AECT-2010-1-0007. Data products have been stored at the Port d'Informaci\'o Cient{\'i}fica (PIC).  Fundiing for this project was partially provided by the Spanish Ministerio de Ciencia e Innovacion (MICINN), projects 200850I176, AYA2009-13936, Consolider-Ingenio CSD2007- 00060, research project 2009-SGR-1398 from Generalitat de Catalunya and the Juan de la Cierva MEC program.\\

{ {\it Author contributions:}  All authors contributed to the development and writing of this paper.  The authorship list is given in three groups:  the lead authors (BG, CH, MA) followed by two alphabetical groups.  The first alphabetical group includes those who are key contributors to both the scientific analysis and the data products.  The second group covers those who have either made a significant contribution to the data products, or to the scientific analysis.}
\end{acknowledgements}
\bibliographystyle{aa} 
\bibliography{references} 
\end{document}